\documentclass[12pt,a4paper]{article}
\usepackage{csquotes}
\usepackage{booktabs}
\usepackage{adjustbox}
\usepackage{placeins}
\usepackage{float}
\usepackage{latexsym}
\usepackage{epsfig,longtable,lscape}
\usepackage[super,comma]{natbib}                                                   
\usepackage[table]{xcolor}
\usepackage{amsmath,amsfonts,euscript} 
\usepackage{amssymb}
\usepackage{verbatim,alltt}
\usepackage{float}
\usepackage{psfrag}
\usepackage{color}
\usepackage{times}
\usepackage{rotating}
\usepackage[normalem]{ulem}
\usepackage[T1]{fontenc}
\usepackage{mathptmx}
\usepackage{bigdelim}
\usepackage{booktabs}

\allowdisplaybreaks[4]
\usepackage{longtable,lscape}
\usepackage{hyperref}

\allowdisplaybreaks[4]
\usepackage{longtable,lscape}
\usepackage{hyperref}
\usepackage{setspace}
\begin{document}

\long\def\comment#1{}  
\def\p{\partial}
\def\ds{\displaystyle}
\def\sc{\scriptscriptstyle}
\def\vdot{\mbox{\large \bf .}}

\newcommand{\Frac}[2]{{\displaystyle #1 \over \displaystyle #2}}
\newcommand{\bfmath}[1]{\mbox{\boldmath$#1$}}
\newcommand{\fs}[2]{\fontsize{#1}{#2}\selectfont}
\newcommand{\revrea}[2]{ \overset{\ds #1}{\underset{\ds #2}{\mbox{\scalebox{2}{$\rightleftharpoons$}}}} }

\title{\textbf{Accurate Prediction of Global Mean Temperature through Data Transformation Techniques} \\
   \large [PRE-PRINT]    
   }

\author{Debdarsan Niyogi\footnote{E-mail:
        debdarsann.iisc.ac.in (corresponding author) ORCID ID: 0000-0002-2376-5482} \\
		J. Srinivasan\footnote{E-mail:
      jayes@iisc.ac.in, Ph:
      +91-80-22933068} \\
  Divecha Centre for Climate Change, \\
  Indian Institute of Science,
  Bangalore 560012, India }
\maketitle

\begin{abstract}
It is important to predict how the Global Mean Temperature (GMT) will evolve in the next few decades. The ability to predict historical data is a necessary first step toward the actual goal of making long-range forecasts. This paper examines the advantage of statistical and simpler Machine Learning (ML) methods instead of directly using complex ML algorithms and Deep Learning Neural Networks (DNN). Often neglected data transformation methods prior to applying different algorithms have been used as a means of improving the predictive accuracy. The GMT time series is treated both as a univariate time series and also cast as a regression problem. Some steps of data transformations were found to be effective. Various simple ML methods did as well or better than the more well-known ones showing merit in trying a large bouquet of algorithms as a first step. Fifty-six algorithms were subject to Box-Cox, Yeo-Johnson, and first-order differencing and compared with the absence of them. Predictions for the annual GMT testing data were better than that published so far, with the lowest RMSE value of 0.02  $^\circ$C. RMSE for five-year mean GMT values for the test data ranged from 0.00002 to 0.00036  $^\circ$C.

\end{abstract}

{\bf{Impact Statement}} \\
Prediction of Global Mean Temperature has been tried to maximize accuracy using non-ANN methods. Literature suggests (Mkridakis et al, 2018) that an overwhelming real-life time series can be better predicted than by sophisticated and complicated Deep Neural Networks. So, a study of fifty-six machine learning algorithms has been tried to find the most accurate one. In the algorithm search, the input data were transformed using various ways and they were found to have a dramatic effect on the accuracies of the algorithms. The accuracy of the prediction is found to be the highest than those in the literature.


\section{Introduction}

The global mean temperature of the earth is 1.1 $^\circ$C above its
pre-industrial value. This rapid increase is a cause for global
concern. Hence most of the countries in the world signed an agreement
in Paris in 2015 to limit global warming to below 2 $^\circ$C
(preferably 1.5 $^\circ$C). Most nations have not committed to a rapid
reduction of emission of greenhouse gases and hence the Global Mean
Temperature (GMT) will continue to increase. When will the GMT reach
the value of 1.5 $^\circ$C above the pre-industrial value? Can this be
forecast using climate models or methods of data analytics?  Climate
models have been used to predict the GMT at 2100 based on various
assumptions about the increase in greenhouse gases. These models were
not very successful, and not accurate for predicting the increase in
GMT in the near future. A data-driven approach to this problem is an
alternative. Further, to enhance confidence in forecasts, it is
necessary as a first step to predict recent trends in GMT. Hence, it
is useful to examine if ML models will provide accurate
predictions of historical record of the global mean temperature.
ML methods range from simple algorithms based on
statistical analysis and simple Machine learning  to
complex  ML models like Deep Learning Neural
Networks (DNN). The more complex the method is, greater is the effort
and expense in computation.  A few publications strongly recommend using a gamut of
simple ML models prior to attempting the more complex
ones. \citet{Makridakis2018} studied 1045 real-life monthly time series used in
the M3 Competition of the International Institute of Forecasters  to
compare the skills of eight traditional statistical models against
popular ML methods.  He found that simpler ML methods showed better
skills than Artificial Neural Networks (ANN).
\citet{Crone2011} studied the data sets used in NN3 Neural Networks
Forecasting competition under different conditions like (a) short and
long series (b) seasonal and non-seasonal (c) short, medium and long
forecast horizons. Quite surprisingly, the simple Theta and ARIMA models
outperformed other algorithms including neural networks. The value and impact of knowledge from statistics on the methods and results
of data science approaches have been pointed out by
\citet{weihs2018data}. 
Hence,  in the present work only statistical and simple ML methods
were used as the first step to predict the GMT for 5, 10 and 15 years and also mean GMT over one and a half decade.

\section{Literature Review}
The literature has centered around a few lines of thought. One is about
analyzing regional temperature data to infer statistical properties. \citet{Lorentzen_2014} examined sea temperature data from Norway with an aim to infer changes
in climate. \citet{Mali_2014} analyzed multifractal characteristics to examine
if there are long term trends in global temperature anomaly. \citet{Silva_2018}
characterized the variations of global sea surface temperature using wavelets.
\citet{Coleman_2022} used an advanced variant of Empirical Mode Decomposition to
extract temperature cyles in global, hemispherical and tropical temperature
anomalies. There have been many other studies of a similar nature which are
aimed at understanding nature of the time series which will eventually
go into forecast. However, predicting or forecasting global mean temperature 
with the yearly data available is also of great interest.

\citet{mulrennan2020modelling} compared one standard statistical model
ARIMA with two ML models: Artificial Neural Network (ANN) and Random
forest. The time series data were on energy consumption in
pharmaceutical manufacturing facility. Considerable effort was spent
in developing the models and they found Random Forest performed the
best. This is indicative that ANN might perform poorly than a much
simpler ML algorithm. It would have been interesting if predictions
were made with simpler ML models, which is proposed
in this work.

The ML literature on GMT focused on developing correlations with
greenhouse gas emissions, results of Global Circulation Model (GCM) and so on. Very few studies
could be found on characterizing or predicting GMT by treating it as a
time series. Partial Least Square Regression (PLR) was used by
\citet{Brown2020} to analyze the GMT time series. Observed globally
gridded Surface Air Temperature (SAT) anomalies were obtained from
four primary data sets. SAT fields were used as predictor variables of
GMT.  PLSR was performed between predictors made up of gridded SAT
fields and predictands of subsequent GMT deviations. Model (called
BC2020) validation was achieved via leave-one-out-cross validation for
years prior to 2000 (which is commonly referred to as hindcast mode)
and through out-of-sample predictions made on the post-2000 data
(which is commonly referred to as forecast mode). The walk-forward
method was adopted where the window size varied from 1 to 4. The
lowest value of RMSE in the forecast was 0.09.  One noteworthy
observation of \citet{Brown2020} is that the mean GCM had a larger
RMSE than the Naïve benchmarks which suggests that the average GCM has
difficulty in predicting GMT deviations.

Monthly data during 1974-2020 from data of 270 air temperature
measuring stations in Turkey were analyzed by \citet{Citakoglu2021}.
Data from 165 stations were used for training and the others for
testing. Month numbers, latitude, longitude, and altitude variables
were used as input data. Performances of four different ML methods (a)
Long Short-Term Memory (LSTM), (b) Support Vector Machine Regression
(SVMR), (c) Gaussian Process Regression (GPR), and (d) Multi-Gene
Genetic Programming (MGGP) were compared. The models were used to
predict maximum, minimum and average temperatures.  A typical value of
RMSE (lowest for the prediction of the average temperature) for all
the four models was 1.28.

\citet{Panda2021} used DNN methods (LSTM, GRU, Bi-LSTM, Bi-GRU,
S-LSTM, S-GRU, SBi-LSTM, SBi-GRU) to model GMT data from Berkley's
Earth between January 1900 and June 2020 with a train test split of
80:20. They compared the predictions with those of the statistical
model SARIMAX. The latter was found to be more accurate with an RMSE
of 0.0838.

\citet{Viola2010} employed nonlinear signal processing tools to
analyze temperature time series data obtained at different locations
in the world from January 1989 to December 2008. The aim was to
establish state-space reconstruction and make predictions. Special
techniques were used to reduce noise contamination. The testing period
was about eight years. The average error is defined as the per cent of
the difference between observed and predicted values over the time
series divided by the observed mean. The error ranged from 0.6 to
5.3\% for the different locations.

\citet{Romilly2005} used univariate time series of GMT to develop a
forecasting model over the short-term horizon (5 to 10 years). The
statistical techniques used include seasonal and non-seasonal unit
root testing, as well as ARIMA and GARCH modelling. The dataset
consists of 1602 monthly observations from 1870:1 to 2003:6 on global
near-surface mean temperature, expressed as differences (or anomalies)
between the actual monthly temperature and the average temperature for
1961–90. Temperature is measured in degrees Celsius, and the average
temperature from 1961–1990 is 14.08 $^\circ$C. The SARMA–GARCH(1,1) M
model gave the best RMSE = 0.111.

The monthly record of absolute surface temperature was modelled by
\citet{Ye2013} using a Deterministic and Stochastic Combined (DSC)
approach, where the deterministic part consists of trend and cyclic
oscillations, and the stochastic part is the remaining pattern
involving SARIMA. Several DSC models were constructed and tested. The
best RMSE obtained was 0.1112.

\citet{Himika2018} used an
ensemble approach for GMT prediction and reached a range of RMSE from 0.65 to 2.41 for different algorithms tried. Three models were picked out for assembling, and the RMSE 0.67 was obtained for the said ensemble model.

There are climate models \cite{neelin2010climate}, like the Global
Circulation Model (GCM), that are physics-based simulations of
real-world phenomena, developed to advance the understanding of
climatic behaviour. Later, results of some of these are compared with
those obtained in this work.

It can be seen that accuracy attained in prediction of historical data
of GMT  is still unsatisfactory.

\section{Exploratory Data Analysis}
\subsection{Data and Characterization}

\begin{figure}[H]
\begin{center}
  \includegraphics[width=0.8\textwidth]{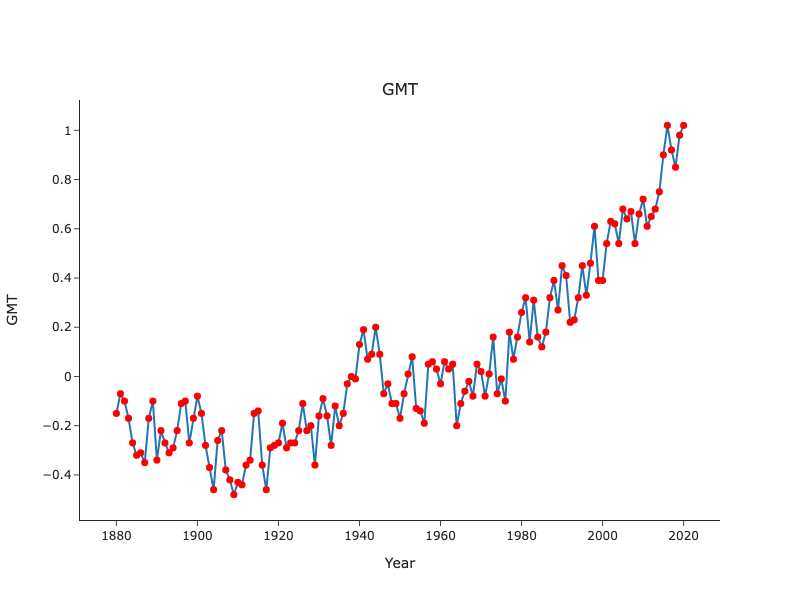}
  \caption{Line plot of GMT data (1880-2020)}
  \label{fig:GMT-LINE}
\end{center}
\end{figure}

\begin{figure}[H]
\begin{center}
  \includegraphics[width=0.8\textwidth]{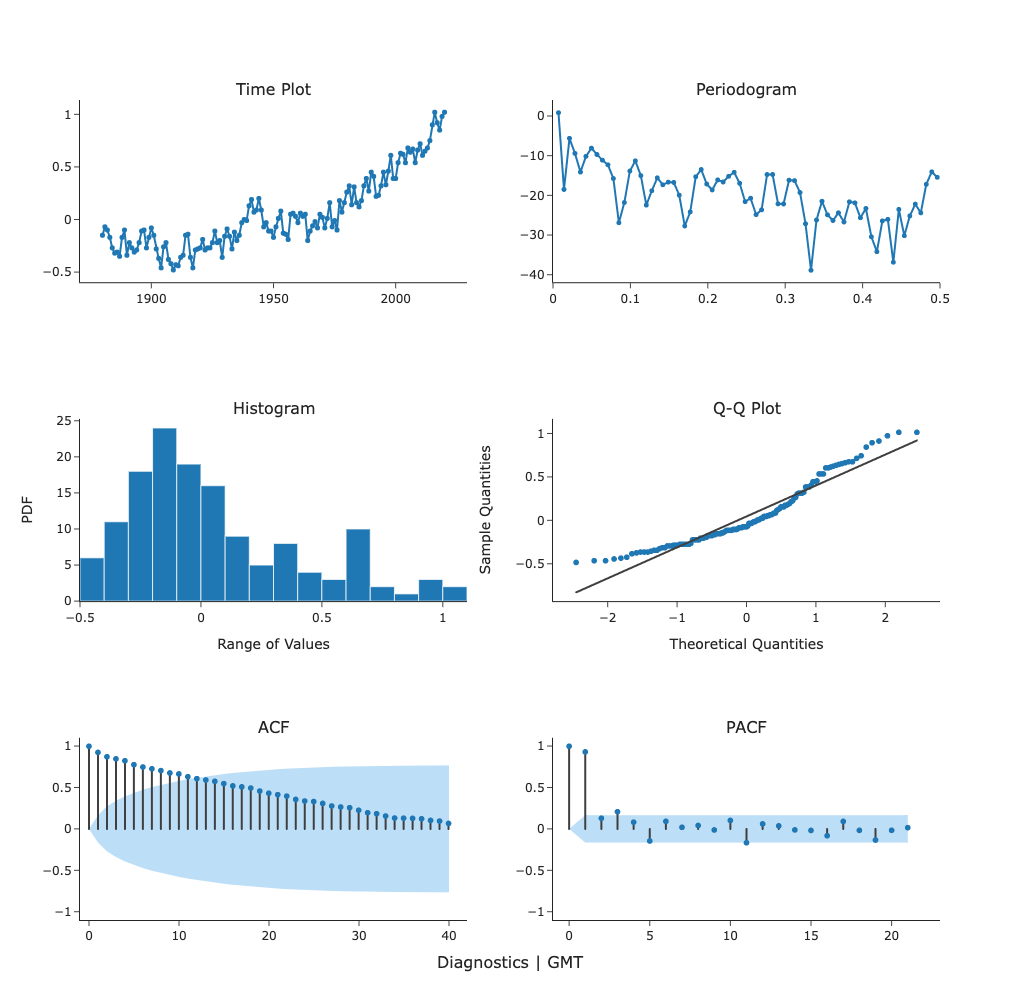}
  \caption{Diagnostic Plots}
  \label{fig:GMT-DIAG}
\end{center}
\end{figure}

\begin{figure}[H]
\begin{center}
  \includegraphics[width=0.8\textwidth]{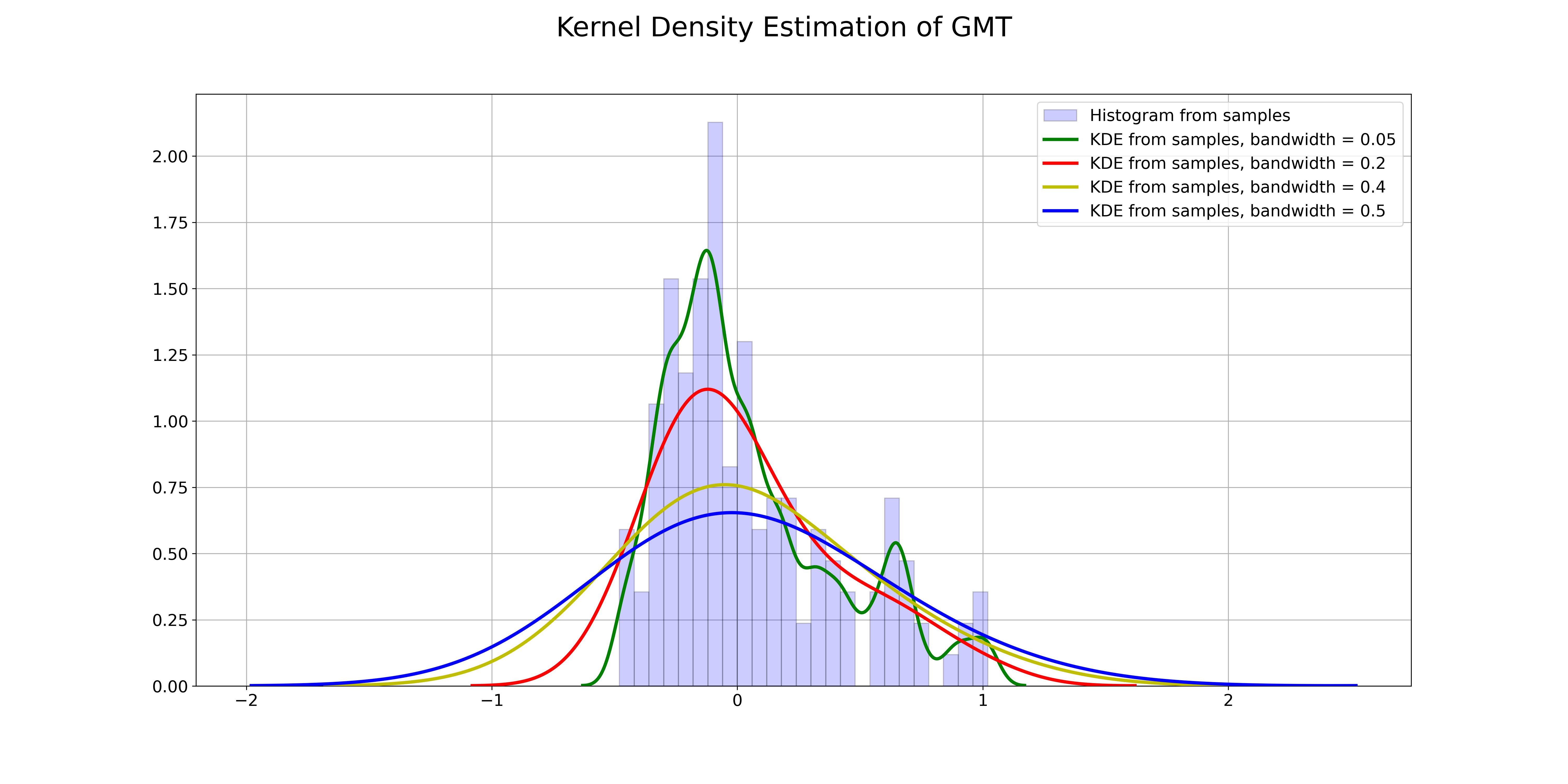}
  \caption{Kernel Density Estimation}
  \label{fig:GMT-KDE}
\end{center}
\end{figure}

The GMT data was obtained from NASA's website from 1880 to 2020
(\url{https://data.giss.nasa.gov/gistemp/}). The data column which lists the
GMT average from January to December, namely \textquote{J-D}, for each
year, has been used in this univariate study. The line plot of GMT
presented in Figure \ref{fig:GMT-LINE} shows an increasing trend. The
descriptive statistics are given in Table \ref{tab:desc-stat}. The
diagnostic plots are shown in Figure \ref{fig:GMT-DIAG} and the Kernel
Density Estimation (KDE) \cite{Wglarczyk2018} using default parameters
is depicted in Figure \ref{fig:GMT-KDE}. It is seen that the KDE
exhibits multi-modality.

\begin{table}[!htbp]
  \centering
  \caption{Descriptive Statistics}
    \begin{tabular}{|l|r|}
    \hline
    \rowcolor[rgb]{ .608,  .761,  .902} \text{Count} & \cellcolor[rgb]{ 1,  1,  1}141 \\
    \hline
    \rowcolor[rgb]{ .608,  .761,  .902} \text{Mean} & \cellcolor[rgb]{ 1,  1,  1}0.0504 \\
    \hline
    \rowcolor[rgb]{ .608,  .761,  .902} \text{Standard Deviation} & \cellcolor[rgb]{ 1,  1,  1}0.3579 \\
    \hline
    \rowcolor[rgb]{ .608,  .761,  .902} \text{Min. Value} & \cellcolor[rgb]{ 1,  1,  1}-0.48 \\
    \hline
    \rowcolor[rgb]{ .608,  .761,  .902} \text{25\%} & \cellcolor[rgb]{ 1,  1,  1}-0.2 \\
    \hline
    \rowcolor[rgb]{ .608,  .761,  .902} \text{50\%} & \cellcolor[rgb]{ 1,  1,  1}-0.07 \\
    \hline
    \rowcolor[rgb]{ .608,  .761,  .902} \text{75\%} & \cellcolor[rgb]{ 1,  1,  1}0.23 \\
    \hline
    \rowcolor[rgb]{ .608,  .761,  .902} \text{Maximum Value} & \cellcolor[rgb]{ 1,  1,  1}1.02 \\
    \hline
    \rowcolor[rgb]{ .608,  .761,  .902} \text{Kurtosis} & \cellcolor[rgb]{ 1,  1,  1}0.0430 \\
    \hline
    \rowcolor[rgb]{ .608,  .761,  .902} \text{Skewness} & \cellcolor[rgb]{ 1,  1,  1}0.9049 \\
    \hline
    \end{tabular}%
  \label{tab:desc-stat}%
\end{table}%

\begin{figure}[H]
\begin{center}
  \includegraphics[width=0.8\textwidth]{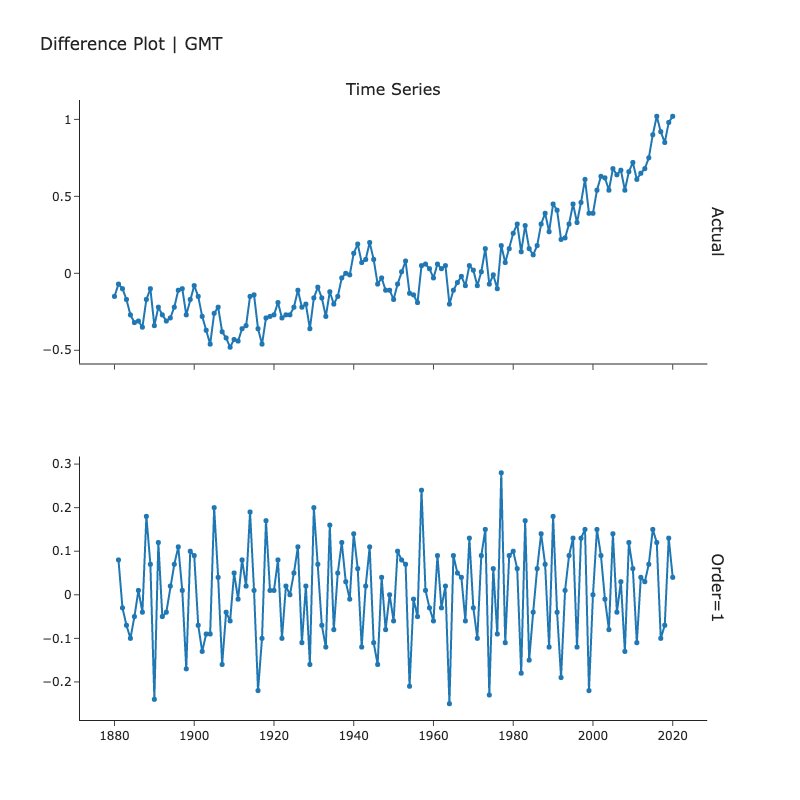}
  \caption{First order difference of GMT data}
  \label{fig:GMT-DIFFERENCED}
\end{center}
\end{figure}

The Normality test showed that the distribution is not Gaussian.  The
Augmented Dickey-Fuller (``ADF'') test and
Kwiatkowski-Phillips-Schmidt-Shin (``KPSS'') test indicate that the
time series is non-stationary.  The time series after first-order
differencing is shown in Figure \ref{fig:GMT-DIFFERENCED}. Both ADF
and KPSS tests are repeated on the first-ordered differenced data and
they confirm the time series is stationary after differencing.

\begin{figure}[H]
\begin{center}
  \includegraphics[width=0.8\textwidth]{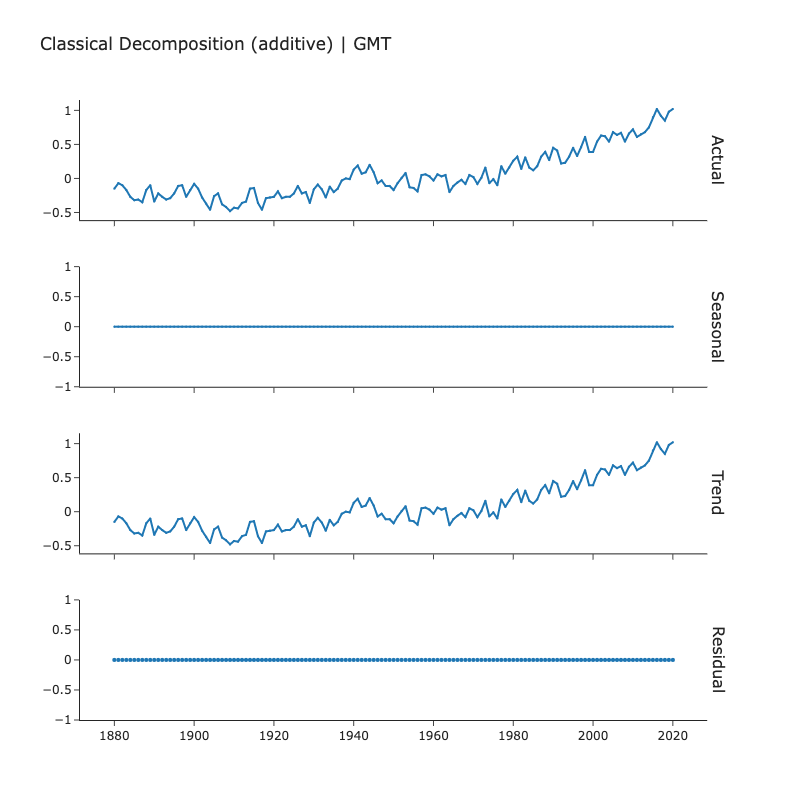}
  \caption{Additive decomposition of GMT time series}
  \label{fig:GMT-ADD-DECOMP}
\end{center}
\end{figure}

The systematic part of the GMT time series was decomposed into the
level, trend and seasonality. Since GMT data contains negative values,
the multiplicative decomposition is not applicable. The upward trend
of GMT can be seen in Figure \ref{fig:GMT-LINE}. There cannot be any
seasonality since GMT data is an annual average, and is confirmed in Figure \ref{fig:GMT-ADD-DECOMP}
where seasonal as well as noise components are zero.
It may be noted that additive decomposition is done by using {\it {statsmodels 0.14.0}}
(\url{https://www.statsmodels.org/dev/generated/statsmodels.tsa.seasonal.seasonal\_decompose.html})
and not to be confused with decomposition done in signal processing (CEEMDAN, EWT and so on).

\subsection{Analysis of Outliers}
\begin{figure}[H]
\begin{center}
  \includegraphics[width=0.8\textwidth]{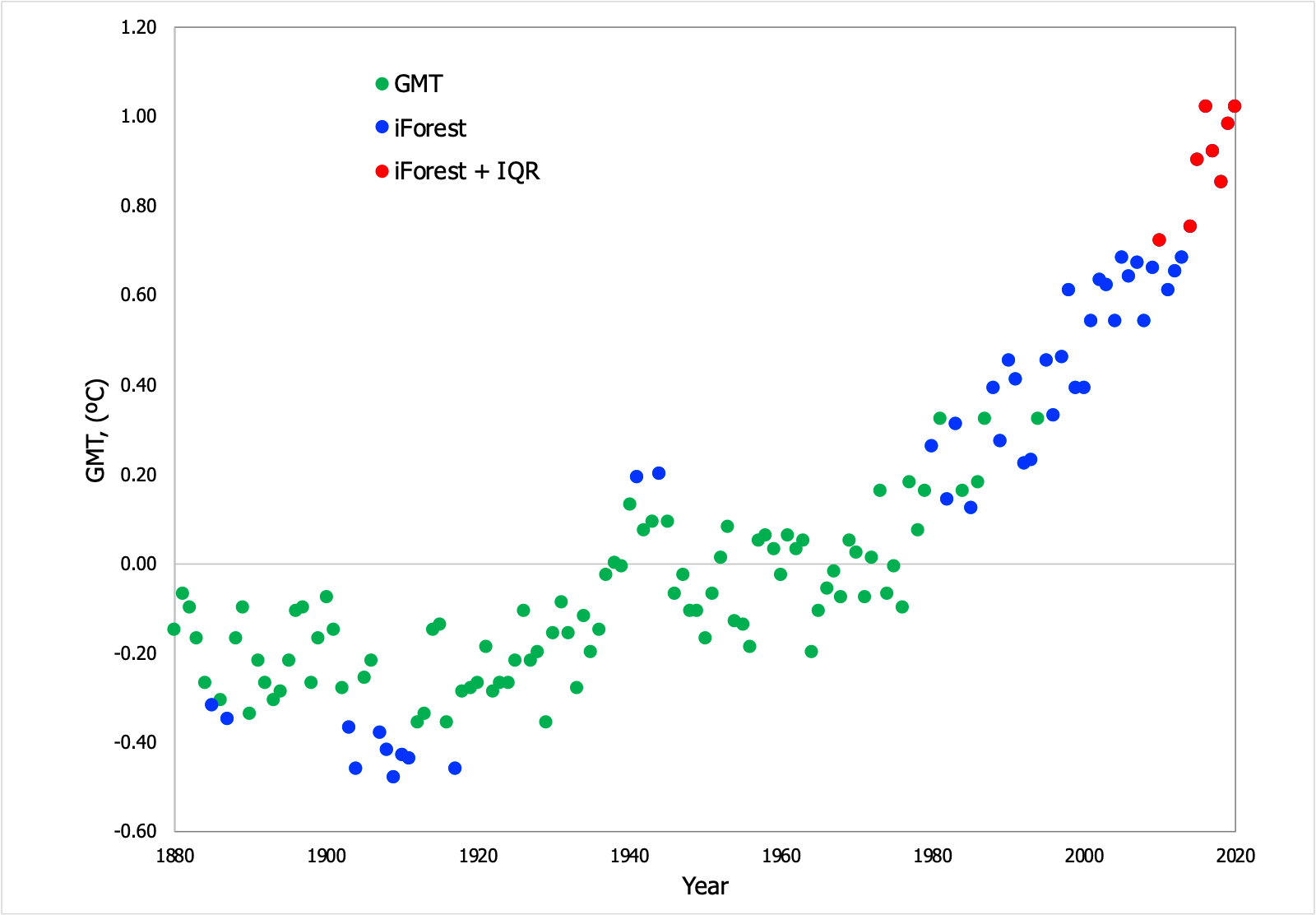}
  \caption{Test for Anomalous Values in GMT Time Series}
  \label{fig:ANO}
\end{center}
\end{figure}

Inter Quartile Range and Isolation Forest methods were used to
identify the outliers and the results are shown on Figure
\ref{fig:ANO}. GMT for the last 8 years are identified as outliers by
both the methods. It is evident that forecasting these by ML methods
will be difficult, as they show departure from the dominant pattern of
variation of GMT time series.

\section{Experiments and Methodology}
A time series predictive problem can be formulated in different ways:
(a) Univariate: where the target variable alone is the dependent one. The time interval must
be uniform and any missing value must be imputed. Conveniently, the
time series of GMT does not have missing values. (b) Multivariate with
exogenous variables: here other independent variables having the same
time axis are combined together with the target variable. (c)
Multivariate formulation using the lagged values as one or more
parallel series, converting the time series to a regression problem,
the lagged series acting as the independent variables. The formulations
(a) and (c) are selected in the present work where the former is
discussed in details, and only the best results for the latter are
presented.

Open-source $Python \ 3.8$ is used as the programming language. To
train, tune and predict a large set of models, $PyCaret$ (an
open-source, low-code machine learning library in Python) is
used. $PyCaret$ is essentially a Python wrapper overarching several ML
libraries and frameworks such as $sktime, sklearn$, regressors like
$XGBoost, LightGBM, CatBoost$, as well as connecting to other
time series packages like $pmdarima$ (for Auto ARIMA), $statsmodel$
(statistical algorithms like ARIMA, ETS) and $Prophet$
(\citet{alipycaret}).  It is a recent development, but gaining
popularity in diverse
fields.~\cite{munjal2020machine,iqbal2021performance,manafiazar2021psxv,ortelli2022optimization}. For the documentation and code examples please refer to {\it PyCaret's} GitHub site: \url{https://pycaret.org/}.

Both the {\it{Time Series}} and {\it{Regression}} modules of
{\it{PyCaret}} have been used for implementing formulations (a) and
(c) respectively and they are discussed under separate heads with the
same names.

A large set of ML algorithms are used in this study (listed in the Table \ref{tab:MODELS} in Appendix) for both time series and regression. As details of these algorithms are available in the books  ~\cite{books0033642,flach2012machine,hackeling2017mastering,brownleemaster}. A few rare algorithms used in {\it PyCaret}, for example ARD, the discussion can be found on the internet by searching the full names. Hence, details of the algorithms are not discussed here.

\subsection{Model Parameters}
\subsubsection{Time Series}
A few parameters are to be defined for carrying out a time series
prediction experiment. The GMT data are available from the year 1880
to 2020, comprising a total of 141 records.  Defining the test size is
mandatory, and test size was varied as 5, 10, and 15 years.  The
training period gets defined accordingly. Other optional techniques
can be used to get a more robust prediction, for example, multiple
train and test splitting. Three-fold splitting which results in three
sets of training and testing data was used here, and the training size
was expended keeping the testing size constant. A schematic diagram
(Figure \ref{fig:cv}) is shown in the Appendix.
 
\subsubsection{Data Transformation}
Three kinds of data preparations (transformations) were tried in this work. This is done to make the data more suitable to the data analytic methods. Advantages of each of these three types of transformed are also described below.
\begin{enumerate} 
\item {\it {Differencing}}: As shown earlier, a first-order differencing made
  the GMT time series stationary, stationarity of time series data
  being an implicit assumption of the applicability of many ML
  algorithms.
\item {\it {Power Transform}}: The Normality test described earlier showed
  that GMT data is non-Gaussian. Much better performance is obtained
  if the distribution can be made closer to Gaussian, as algorithms
  like linear regression and logistic regression explicitly assume the
  variables have a Gaussian distribution.  Although this requirement
  is not stringent for non-linear algorithms, they often perform
  better when variables have a Gaussian distribution having less
  skew. A power transform of the raw data is expected to have the
  desired effect. Two popular approaches for such power transforms
  have been tried. They are (a) Box-Cox and (b) Yeo-Johnson
  Transforms. Details of these are presented in the Appendix.
\item {\it {Scaling}}: Standardization makes the data have zero-mean (when
  subtracting the mean in the numerator) and
  unit-variance.~\cite{han2011data}

$$y' = \frac{x - \bar{y}}{\sigma}$$

where $y$ is the original feature vector, $\bar{y}=\text{average}(y)$ is the mean of that feature vector, and $\sigma$ is its standard deviation. Scaling is applied regardless of the usage of differencing and power transforms. \\
\end{enumerate}

The problem of choosing a transform or a
sequence of transforms, for that matter, which makes the data more
suitable to the models often defies science, and borders upon
art.~\cite{brownlee2020data,kuhn2019feature} To complicate the matter
further, different ML algorithms have different requirements. It is
worth mentioning that sometimes the raw data can be found to be more
effective than the power transformed data, too. In this study, along
with using the raw data, the data preparation methods are
applied, to test their efficacies. This is also shown that while data preparation helped for some algorithms, for others it did not have any significant effect. In order to reap the maximum benefit, data preparation is to be considered as a potential way of improving the results.  Thus, supported by our calculations, it is recommended that data preparation be carried out.

Not many studies include the train-test split as a parameter affecting
the model's skill. \citet{Medar2017} showed that by keeping the test
size constant when the length of the training data is varied, the
skill of the model's prediction also varies. For a larger time series,
this effect may not be observed. The GMT time series is quite short
and hence train-test split was also considered as a parameter in this
study as described above.

A total of six combinations are thus possible for data preparation,
two for whether differencing is done or not multiplied by three for
power transforms (two power transforms plus one for the absence of
them). When the number of folds, fold strategy, tuning iteration and
search method are kept constant, as test sizes are varied from 5, 10,
and 15, combined with the data preparation strategy, a total of 18
experimental set-ups are possible. The modelling parameters are
summarized in Table \ref{tab:EXP}.

\begin{table}[htbp]
  \centering
  \caption{Parameters in Models}
    \begin{tabular}{|l|p{9.75em}|p{3.25em}|p{13.915em}|}
    \hline
    \hline
    \rowcolor[rgb]{ .741,  .843,  .933} \multicolumn{1}{|p{6.335em}|}{\textbf{Context}} & \textbf{Parameter} & \textbf{Symbol} & \textbf{Values} \\
    \hline
    \hline
    \multicolumn{1}{|p{6.335em}|}{Data Preparation} & Differencing order & $D_n$    & $n$ = 0, 1 \\
    \hline
          & Power Transform & $PT$ & $PT = NO$ (No Transform), $BC$ (Box-Cox), $YJ$ (Yeo-Johjnson) \\
    \hline
    \multicolumn{1}{|p{6.335em}|}{Train-Test Split} & Test Size & $Tm$
    & $m$ = 5, 10, and 15 \\
    \hline
    \multicolumn{1}{|p{6.335em}|}{Scaling} & Type of Scaling & \multicolumn{1}{l|}{} & Standard \\
    \hline
    \multicolumn{1}{|p{6.335em}|}{Train/Tune} & Fold  & $F_l$    & $l$ = 3 \\
    \hline
          & Fold Strategy & $F_s$    & $s$ = Expanding \\
    \hline
          & Tuning Iterations & $I_u$    & $u$ = 20, 50 \\
    \hline
          & Search Method & $H_v$    & $v$ = Random \\
    \hline
    \end{tabular}%
  \label{tab:EXP}%
\end{table}%

For each of the mentioned 18 combinations in Table \ref{tab:EXP} all
the time series models available in $PyCaret$ listed in Table
\ref{tab:MODELS} of Appendix were trained, tuned and predictions
made. The predicted data are inverse transformed to get back the data
in the original scale.

It is customary to define and evaluate a baseline model (Naïve
Forecaster in the Table \ref{tab:MODELS}). The simplest Naïve model
does not do any computation, it simply returns the last observed
value. As algorithms, scoring less than Naïve are to be discarded, it
was decided to adopt a slightly more skilful type
\cite{hyndman2018forecasting} of the Naïve models. It makes variable
forecasts which increase or decrease over time, where the amount of
change over time (called the \textquote{drift}) is set to be the
average change seen in the historical data. Thus the forecast for time
$t + h$ is represented as:

$$\hat{y}_{t+h|t} = y_{t} + \frac{h}{t-1}\sum_{t=2}^T (y_{t}-y_{t-1}) = y_{t} + h \left( \frac{y_{t} -y_{1}}{t-1}\right)$$

\subsubsection{Regression}
Machine learning methods that use regression techniques can also be
applied to time series once the latter is converted to a supervised
learning problem, i.e., the time series data is manipulated in such a
way that we have a set of input variables (independent), used as
predictors to predict the dependent variable(s). A window size (or
lag) value ($W$) is defined to achieve this such that $GMT(t-W),
GMT(t-W+1), ..., GMT(t-1)$ are used as predictor-variables to predict
the dependent variable $GMT(t)$, where $t$ denotes time. Therefore,
for the regression problem, $W$ forms another parameter, added to the
parameters described in Table \ref{tab:EXP}, taking on integer values
1, 2, 3, ... and so on. The regression algorithms available in
$PyCaret$ are listed in Table \ref{tab:MODELS} of Appendix. The discussion about the algorithms can be found in the books mentioned before (at the end of section {\it {Experiments and Methodology}}.

\subsection{Model Evaluation}Mean Absolute Error (MAE), Mean Squared
Error (MSE), and Root-mean Squared Error (RMSE), Mean Absolute Percent
Error (MAPE), and Normalised Root-mean squared Error (NRMSE) are
commonly used. RMSE was used in this work. Thus, the RMSE of annual
predictions of GMT during the test period was one metric that was used
to evaluate the efficacy of the models. While the annual variation of
GMT is important, as mentioned earlier, trends in the variation of GMT
averaged over a time interval also have acquired importance as it
gives an indication of long-range trends. Decadal mean has been
commonly employed in climate literature. However, given the small size
of data, only two predictions would be possible if a decade is chosen
as the averaging period even for a testing size of 20 years. Hence an
averaging period of five years has been chosen in this work. The
decadal mean is easier to predict more accurately than annual
variation. Choice of a five-year averaging period can give an
indication of what was observed about the predictability of decadal
mean applies to five-year average as well.

If the test size is $m$, the actual GMT data can be represented by $y
= [y_1, y_2, ..., y_m]$. Given $y$ the model predicts $\hat{y}$:
$$\hat{y} = [\hat{y_1}, \hat{y_2}, ..., \hat{y_m}]$$
RMSE is given by
$$RMSE = \sqrt {\frac{1}{m} \sum_{i=1}^{m} (y_{i} - \hat{y_{i}})^2}$$ 

When two mean values, that is observed mean and the mean calculated from the annual prediction data ($\bar{\hat{y}}$)
$$ {\displaystyle \bar{\hat{y}}={\frac {1}{m}}\sum _{i=1}^{m}\hat{y_{i}}} $$

are considered, $m$ becomes unity and the $RMSE (of Mean)$ is given by:

$${\displaystyle RMSE(of \ mean) = abs(\bar{y} - \bar{\hat{y}})}$$

\section{Results \& Discussion}

\subsection{Time Series}
All the results presented here were obtained with the number of folds equal to 3, expanding training window, random hyperparameter search method and tuning iteration of 50. Tables \ref{tab:RMSE-D0} and \ref{tab:RMSE-D1} summarizing all the results are placed in the Appendix, and they show the performances of all the algorithms (thirty and one Naïve) for all the six data preparation methods.  It is to be noted that GRM and PAR are excluded because both of them scored very high RMSE values.  The parameters are indexed by the choice made for differencing, power transformation, and test sizes. For example, D1-YJ-T10 means first-order differencing is used (D1), power transform is Yeo-Johnson (YJ), and test size was 10 (T10).  In the following discussion, a $run$ signifies a particular data preparation and test size. As mentioned earlier, the total number of runs is 18 for each of the algorithms considered.  The results obtained are discussed as follows. A broad comparison of the effect of data preparation on predictions is presented. It is followed by a comparing RMSE of predictions made by various algorithms.  The effect of test size is then discussed.  The results of one  model are used to study the agreement of a 5-year
sliding mean between the predicted and the observed values. The accuracy of this prediction is then compared with that of the annual
prediction for ADA. A comparison of 5, 10 and 15 years of predicted versus actual values is also presented for the algorithm KNN.

\subsubsection{Choice of Algorithms for Discussion}
All the thirty time series algorithms and one Naïve, available in {\it PyCaret} have been evaluated for all the runs. Since the obtained result set is huge, a few representative algorithms are discussed here. The choices are based on two aspects: (a) simplicity (easy to understand and implement) and (b) representative of different classes of algorithms. For this purpose,  a broad classification of algorithms was done comprising of classical statistical method, regression,  distance/instance-based, tree-based and  ensemble algorithms:
\begin{itemize}
\item[(a)] Classical statistical algorithm: This class of algorithms employ statistical methods like auto-regression, moving average, exponential smoothing and so on. ARIMA (ARI), is chosen which integrates the autoregressive and moving average terms and is widely used. 
\item[(b)] Regression: Simple linear regression (LIN). 
\item[(c)] Distance/Instance based:  K-Nearest Neighbour (KNN).
\item[(d)] Tree based: Decision tree (DTR). 
\item[(e)] Ensemble: Between \textquote{bagging} and \textquote{boosting} ensemble methods, \textquote{boosting} is chosen, where the latter boosts the skill by ensembling weak-learners, such as DTR. AdaBoost (ADA) is chosen to be specific. 
\end{itemize}

\subsubsection{Effect of Data Transformation}
The algorithms applied to the time series often assume or require that the time series be stationary. As discussed earlier, the GMT time series is not stationary and as can be seen from Figure \ref{fig:GMT-DIFFERENCED}, it was made stationary by first-order differencing the time series. Power transforms help remove skewness of the data which may render a time series more suitable to ML algorithms. The input data to the algorithms is thus a joint effect of differencing and power transforms. 

\subsubsection{Effect of Differencing}
\begin{figure}[H]
\begin{center}
  \includegraphics[width=1\textwidth]{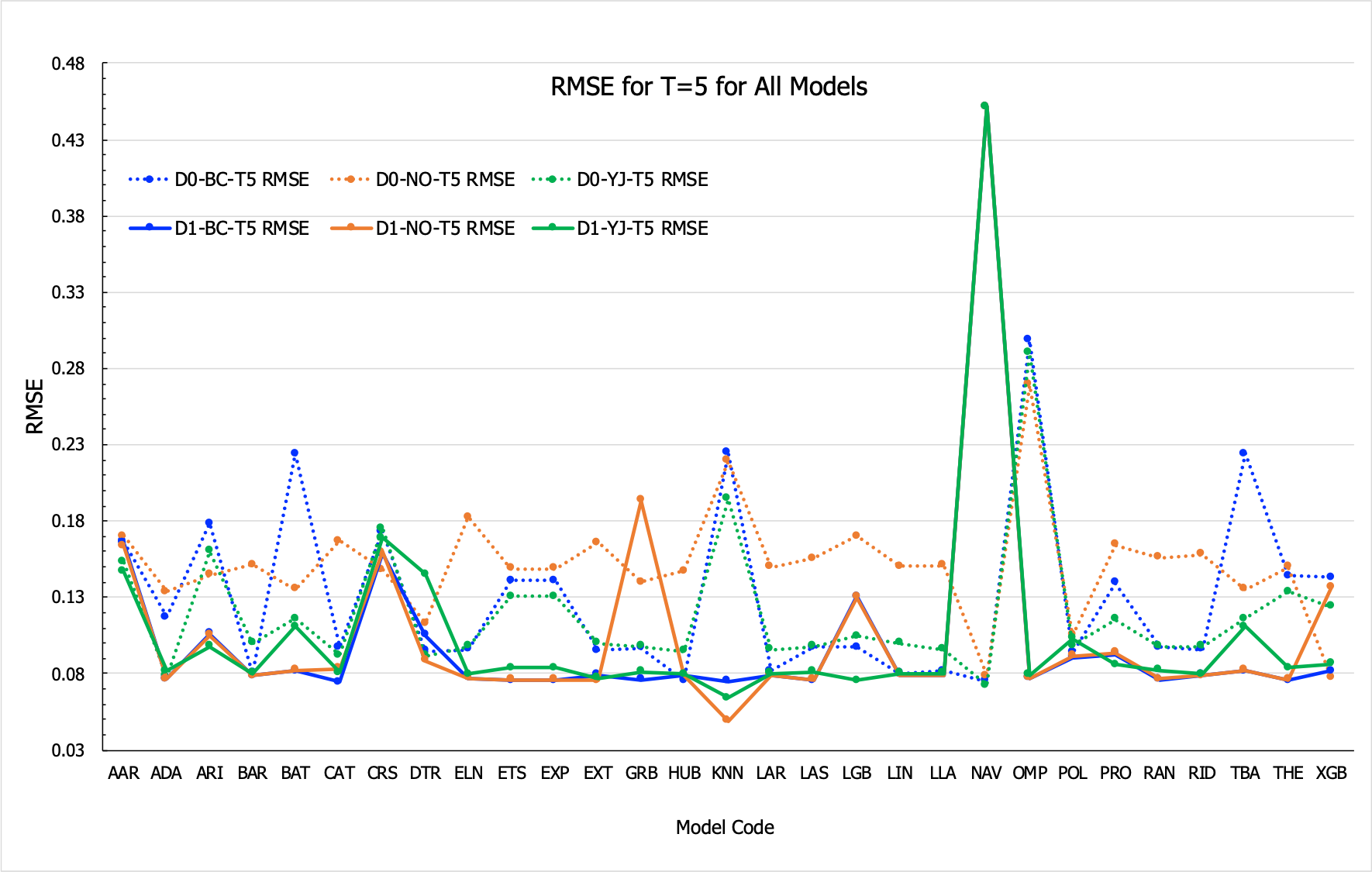}
  \caption{Effect of Data Preparation: All Algorithms, T = 5}
  \label{fig:RMSE-ALL-T5}
\end{center}
\end{figure}

\begin{figure}[H]
\begin{center}
  \includegraphics[width=0.8\textwidth]{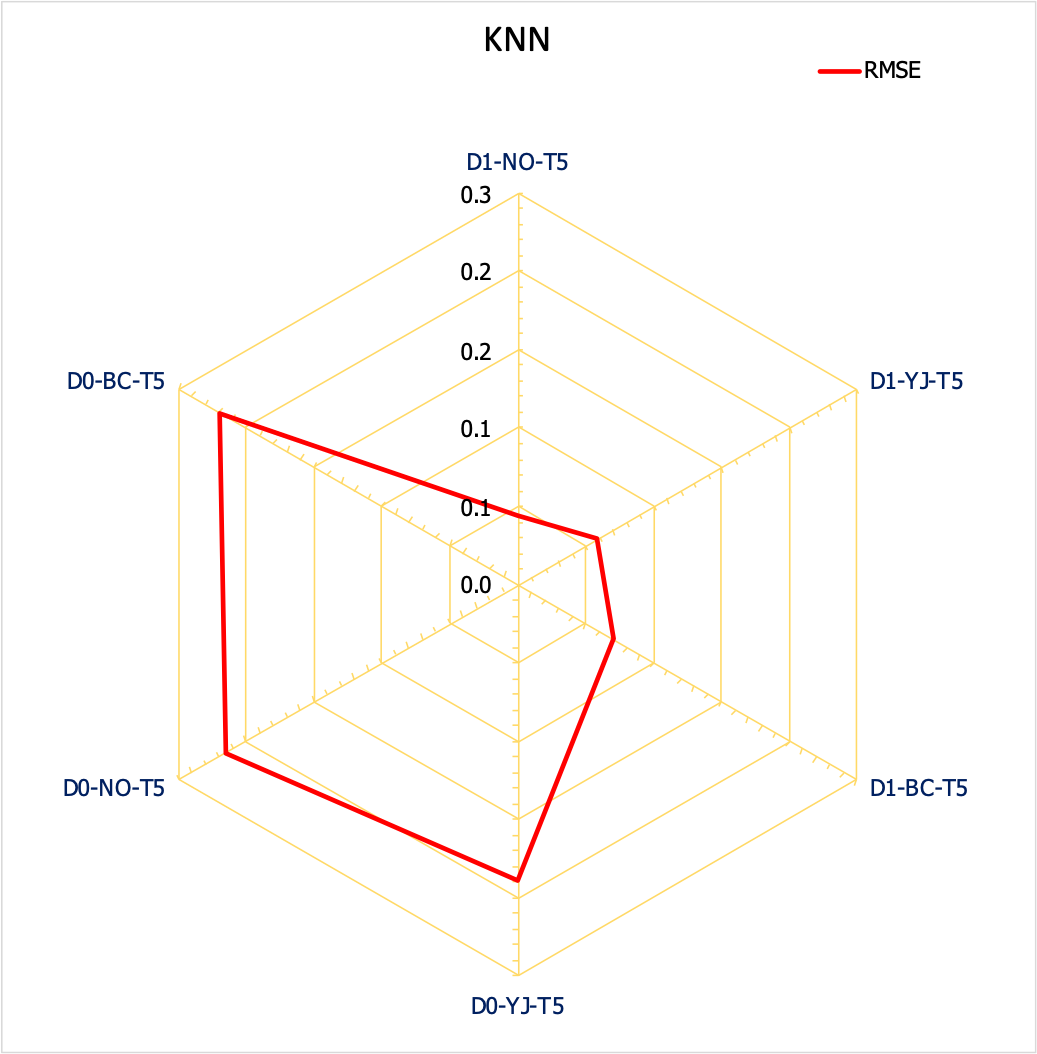}
  \caption{Effect of Data Preparation: KNN}
  \label{fig:KNN-T5}
\end{center}
\end{figure}

Figure \ref{fig:RMSE-ALL-T5} shows the effect of differencing for all the models for a
test size of 5. 
It can be seen that by and large the dotted lines representing no differencing (D0) lie above the solid lines representing first-order differencing (D1). Thus,  first-order differencing gives better results than no differencing. This behavior was also observed for the vast majority of the algorithms for the other test sizes as well, and this can be seen from the tables of detailed results presented in the Appendix. However, there are exceptions. For example, D0-BC-T5 does far better than D1-BC-T5 for NAV. However, these are few and constitute a minor fraction. It should also be mentioned that for D0-BC-T5, and D0-YJ-T5, NAV has the lowest RMSE of 0.067 among the all available algorithms, a surprising result indeed. Moreover,  no other algorithm in the whole set with D0 could reach the lowest RMSE. This also signifies the importance of first-order differencing.  This is reinforced by taking KNN as an example and is pictorially represented in Figure \ref{fig:KNN-T5}.
Here, it can be seen that  D1 does perform better than D0 when power transform and test size are kept the same.

Therefore, only results where differencing was implemented are discussed and the results where data has not been differenced (D0) are not considered for further discussion.

\subsubsection{Effect of Power Transform}
\begin{figure}[H]
\begin{center}
  \includegraphics[width=1\textwidth]{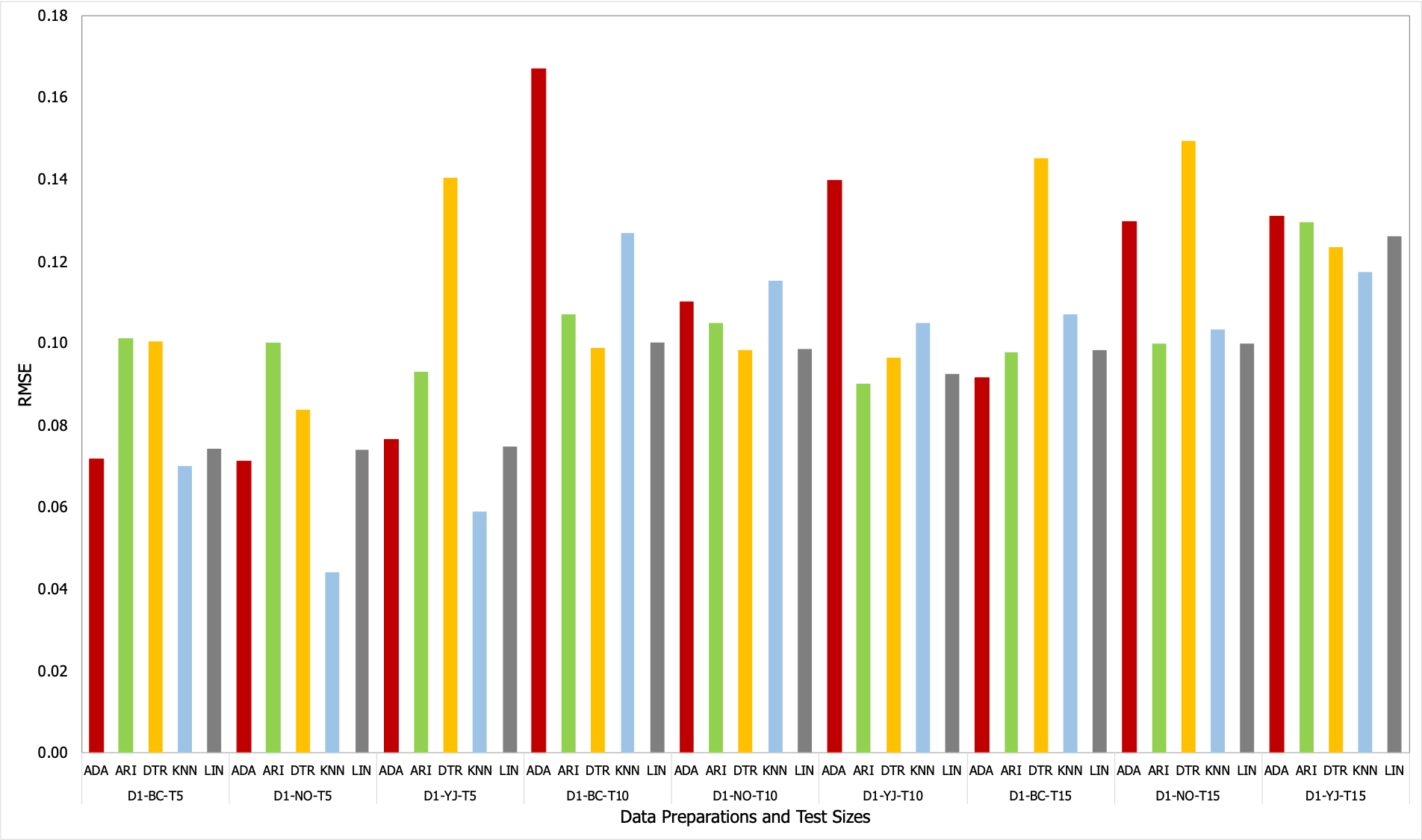}
  \caption{Joint Effect of Differencing, Power Transforms and Test Sizes for 5 Selected Models}
  \label{fig:Data-Prep}
\end{center}
\end{figure}

\begin{figure}[H]
\begin{center}
  \includegraphics[width=0.8\textwidth]{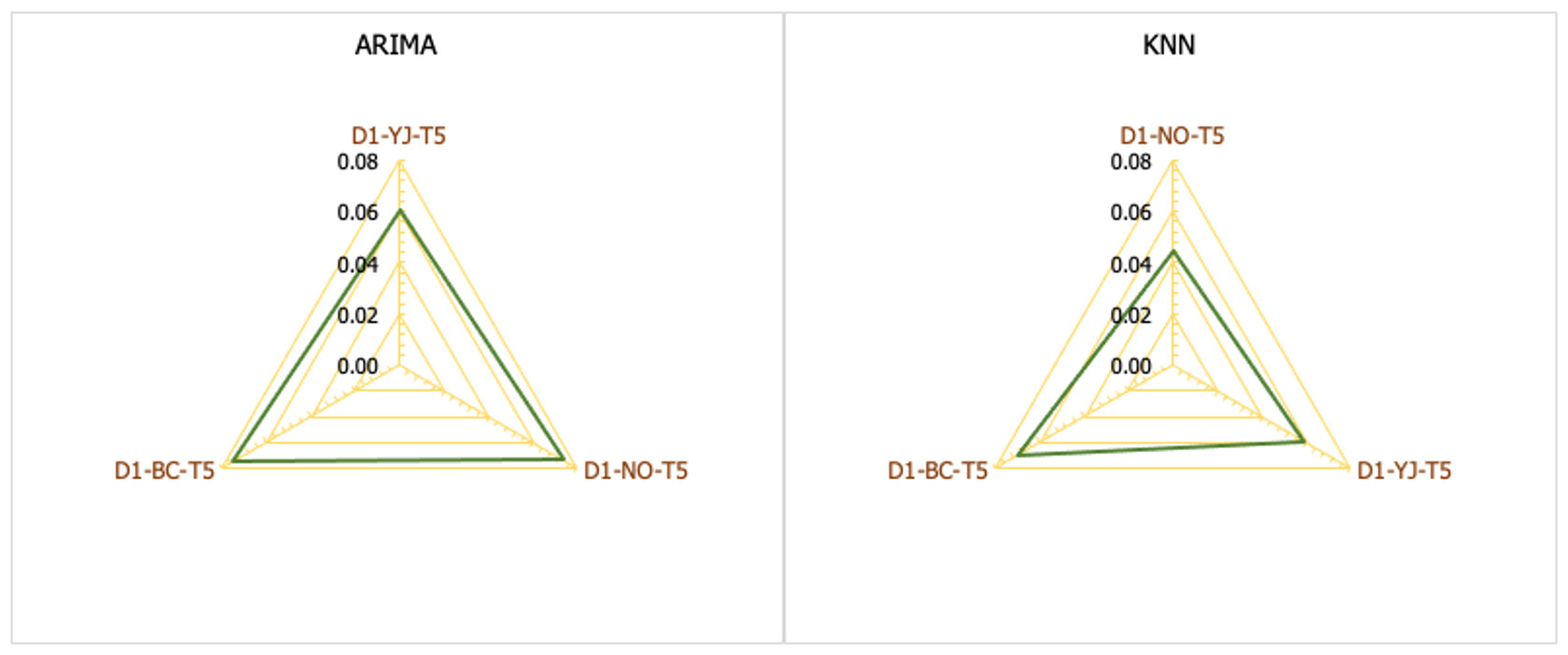}
  \caption{Effect of Power Transform for ARI and KNN (T=5)}
  \label{fig:Effect of PT}
\end{center}
\end{figure}

Figure \ref{fig:Data-Prep} compares the effect of no transform or Box-Cox (BC) or Yeo-Johnson (YJ) transformed data, and for test sizes of 5, 10, and 15. Results for the five selected models are shown as a bar chart.  As mentioned earlier, the comparison is shown only for D1. Each panel begins with red representing ADA and ends with grey representing LIN. Panels corresponding to BC, NO (no transform), and YJ for a constant test size are placed next to each other. The arrangement is repeated for other test sizes. The test size increases from left to right of the figure.  No discernible trends are seen. This is because the effect of test size is mixed with the effect of power transform, and the latter is different on different algorithms. To illustrate this, Figure \ref{fig:Effect of PT} shows the effect of power transform on ARI and KNN. It is seen that for ARI, the order is YJ, NO and BC, whereas for KNN it is NO, YJ and BC.

\subsubsection{Effect of Test Size}An important parameter studied was the train-test size split. Figure \ref{fig:Data-Prep} shows the effect of test sizes too. The effect of test size can be thought of as twofold. Firstly,  the training data available for the algorithm is less as the test size is increased. As the GMT time series is constituted of only 141 values, increasing the test size from 5 to 15 is a significant change in the experimental set-up. Secondly, the primary assumption of all ML algorithms is that the underlying patterns learnt in the training phase are also present in the test data. These two have contradictory effects. Here, no clear trend emerges, however. The test size of 5, and 10 give lower RMSE than 15. This is expected as one anticipates better results as the size of the training set increases. However, the total size of the data set itself is not large,  and the train-test split may not come into play significantly. Also, as discussed earlier, the leading eight data were found to be outliers, which complicates the issue further.

\subsubsection{Comparison of Predictions vs Observations}
\begin{figure}[H]
\begin{center}
  \includegraphics[width=0.8\textwidth]{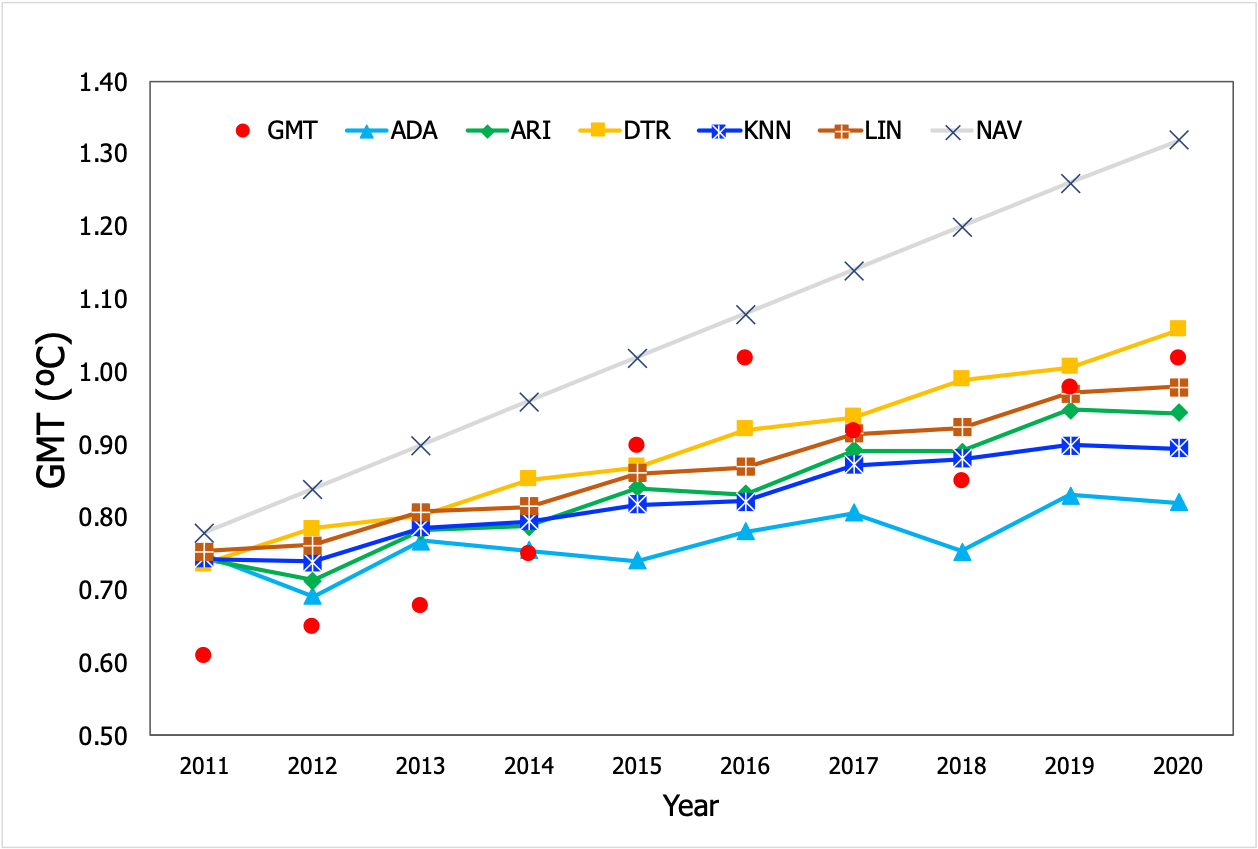}
  \caption{Observed vs Prediction for the Chosen Algorithms and Naïve (T = 10)}
  \label{fig:5-Models+Nav}
\end{center}
\end{figure}

Comparison of predictions with observations of GMT for the five representative models chosen for discussion is shown in  Figure \ref{fig:5-Models+Nav} for the test size of ten (T = 10). {\it It is to be noted that since the test data has been set aside and not used in the training in any way, the predictions for the test data are the closest approximation to forecasts, for both the time series methods and regression formulation}. Display in a graph for all the models will be crowded, and a table will be equally unwieldy. However, observed trends and hence comments for the other models are similar to what is presented for the representative models. The observed values of GMT are plotted against the predicted values in Figure \ref{fig:5-Models+Nav}.  
The lowest RMSE obtained for ARI is 0.090 in this set.  

\begin{figure}[H]
\begin{center}
  \includegraphics[width=0.8\textwidth]{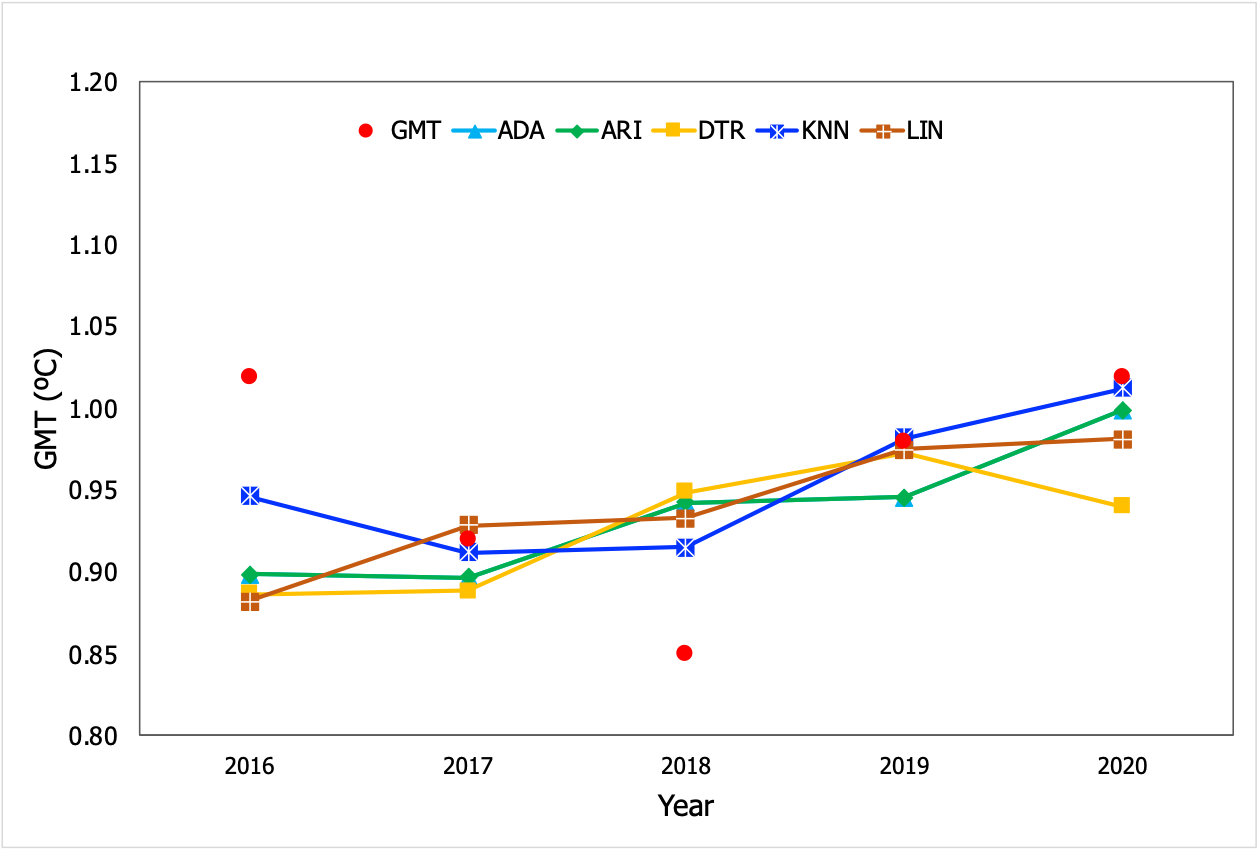}
  \caption{Observed vs Prediction for the Chosen Algorithms (T = 5)}
  \label{fig:5-Models+Nav-T=5}
\end{center}
\end{figure}

Figure \ref{fig:5-Models+Nav-T=5} shows the observed vs prediction for T = 5. For this set, the lowest RMSE (0.044) obtained is by using KNN. This is also the lowest RMSE observed for all the algorithms, all test sizes and data preparation methods.

This value is comparable but lower than what is quoted by \citet{Brown2020} and for predictions by GCM. RMSE values obtained can be considered good since it is comparable to but less than those obtained in GCM models. The results can be considered satisfactory since only the simple ML methods were employed.  Further, this may be considered as a benchmark that must be crossed by the more complicated ML models.

\subsubsection{RMSE of Averaged Values}
\begin{figure}[H] 
\begin{center}
  \includegraphics[width=0.8\textwidth]{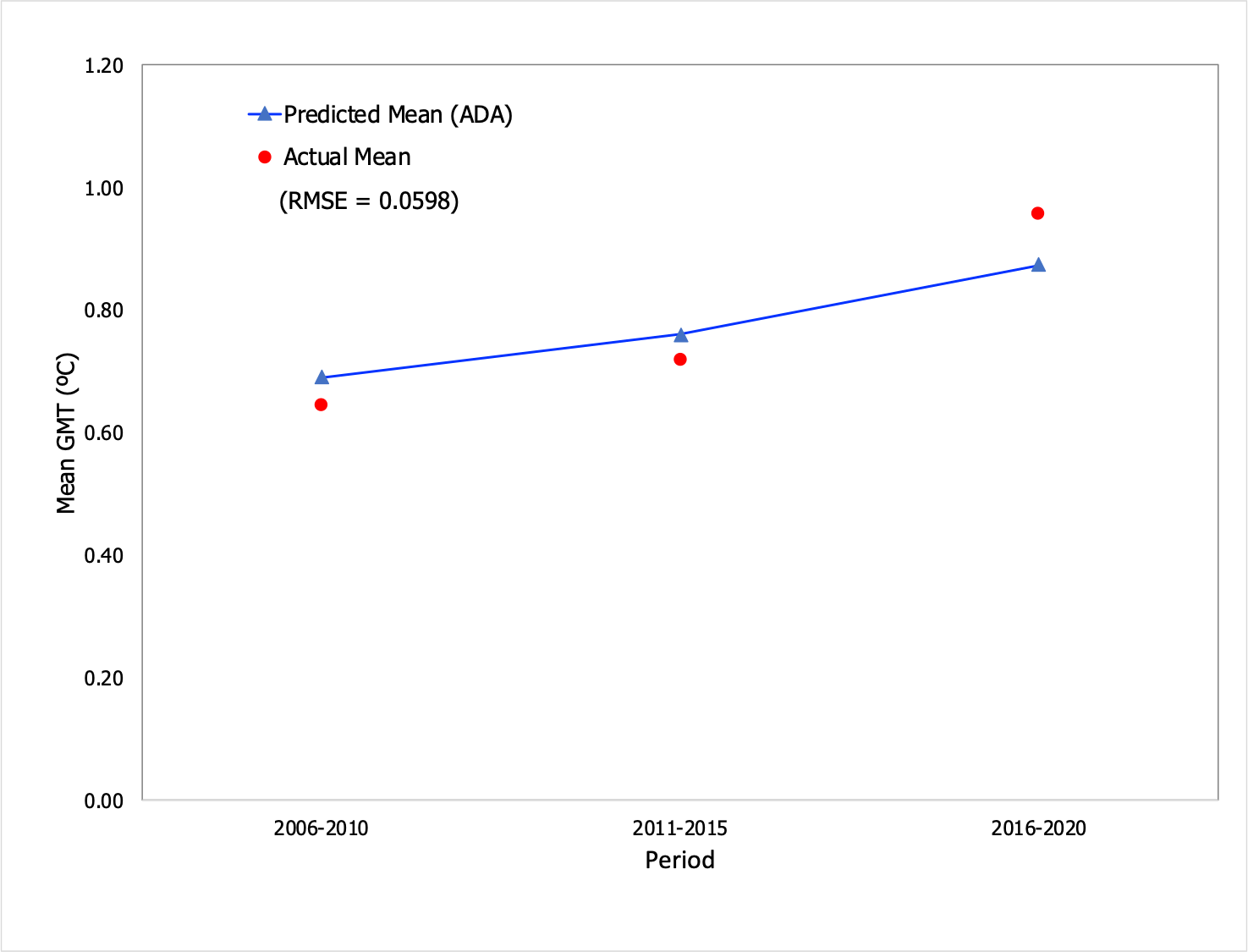}
  \caption{5-year Sliding Mean Observed vs Predicted}
  \label{fig:5-YR-SLIDING-MEAN-T=15-ADA}
\end{center}
\end{figure}

The usefulness of GMT data averaged over five-year periods (referred to here as sliding mean) was mentioned earlier.  This is demonstrated
with a test size of 15 years and with the results of the AdaBoost algorithm (ADA), which was found to be the top performer. The calculation details are as follows. The period is 2006-2020 (15 years). The first mean is one-fifth of the sum of values for 2006 to 2010. The second and the third are for 2011-2015 and 2016-2020. The same applies to computed and observed values. So, three means are calculated from the predicted annual values and another three are actual means calculated from the observed data. RMSE is calculated using these three pairs of mean values.  Predictions of the 5-year sliding mean of GMT made by the ADA are compared with observations in Figure \ref{fig:5-YR-SLIDING-MEAN-T=15-ADA} for each 5-year block. An RMSE of 0.059 was obtained.
This confirms that values averaged over five-year blocks can be predicted with simple ML methods with great accuracy.

\begin{figure}[H]
\begin{center}
  \includegraphics[width=0.8\textwidth]{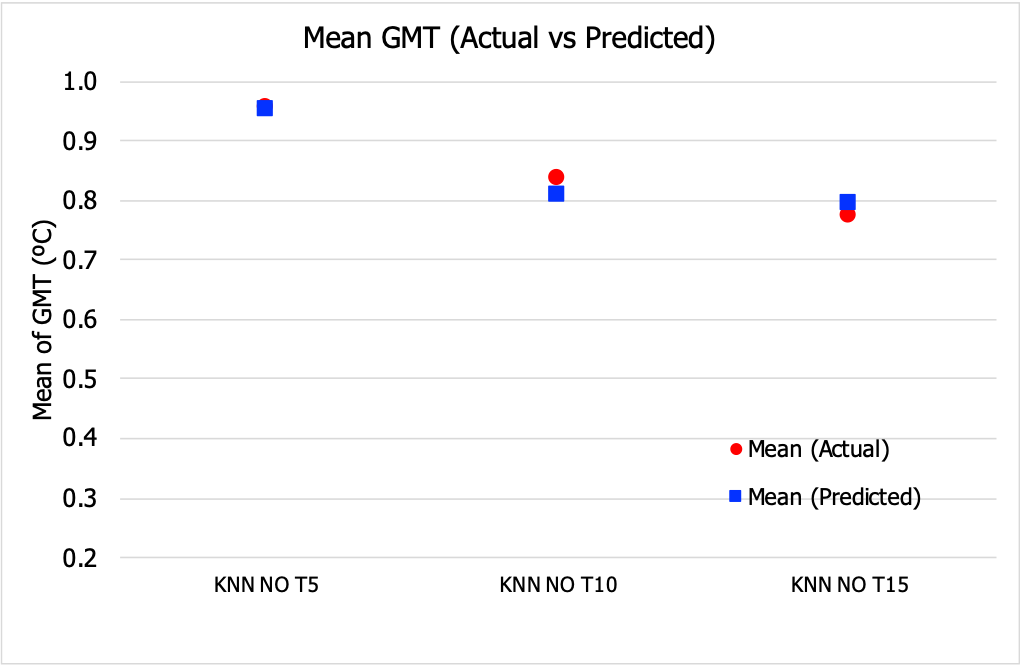}
  \caption{Comparison of 5, 10 and 15 Years Observed and Predicted Means for KNN}
  \label{fig:KNN-Mean-5-15}
\end{center}
\end{figure}

\begin{figure}[H]
\begin{center}
  \includegraphics[width=0.8\textwidth]{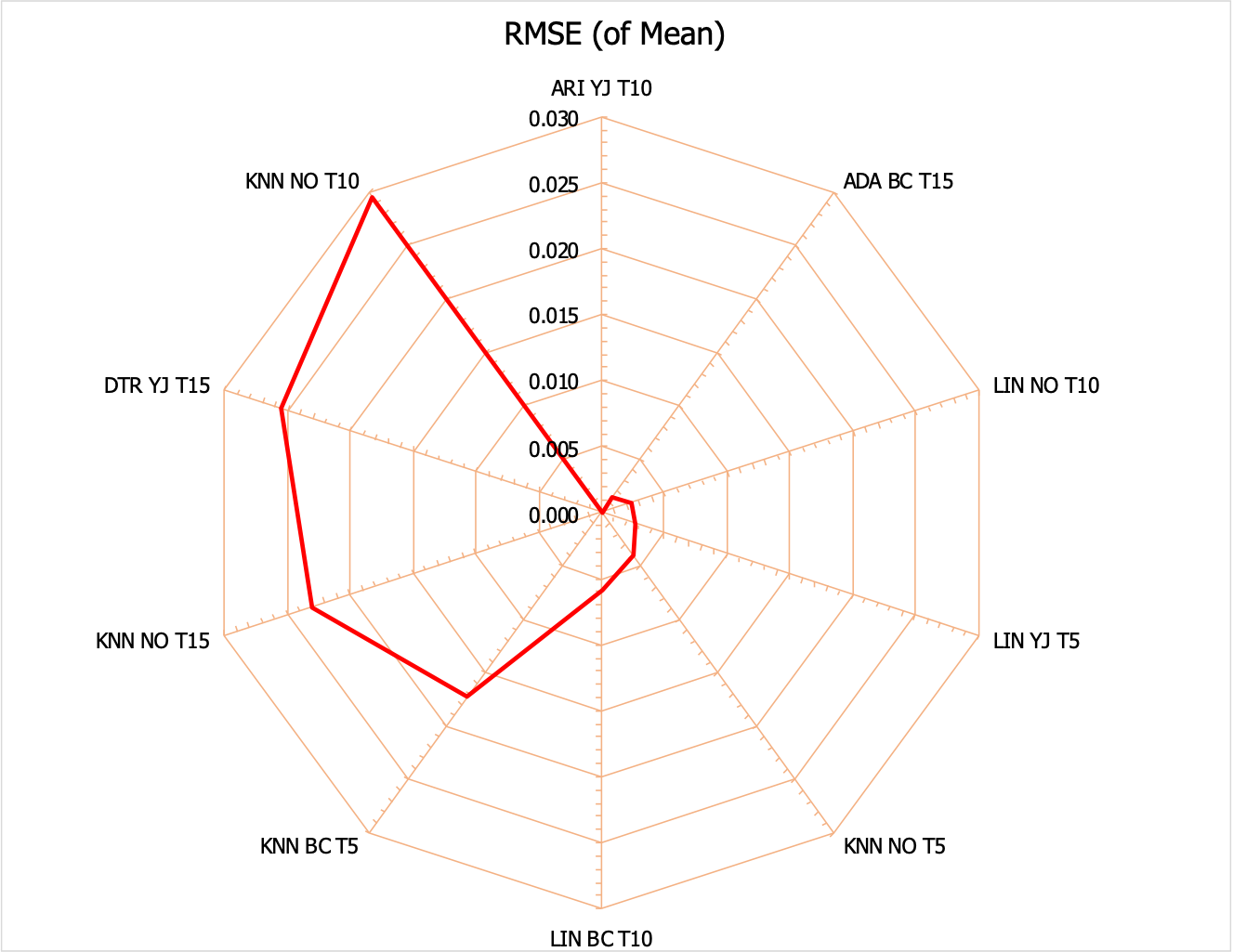}
  \caption{Variation of RMSE (of Mean)}
  \label{fig:RMSE (of MEAN)}
\end{center}
\end{figure}

Predictions of the mean against observations also show a similar trend for other test sizes. The comparison is presented for KNN as an
example. Figure \ref{fig:KNN-Mean-5-15} presents the comparison of observed mean GMT over 5, 10 and 15 years with that of predicted (with RMSE 0.004, 0.029, and 0.023 respectively). Figure \ref{fig:RMSE (of MEAN)} shows how the RMSE (of Mean) varies with other algorithms, data preparation and test sizes. It can be seen the RMSE (of Mean) values here are significantly lower than the RMSE of annual variation of GMT.

\subsubsection{Comparing Models Across All Algorithms All Scenarios}
\begin{figure}[H]
\begin{center}
  \includegraphics[width=0.8\textwidth]{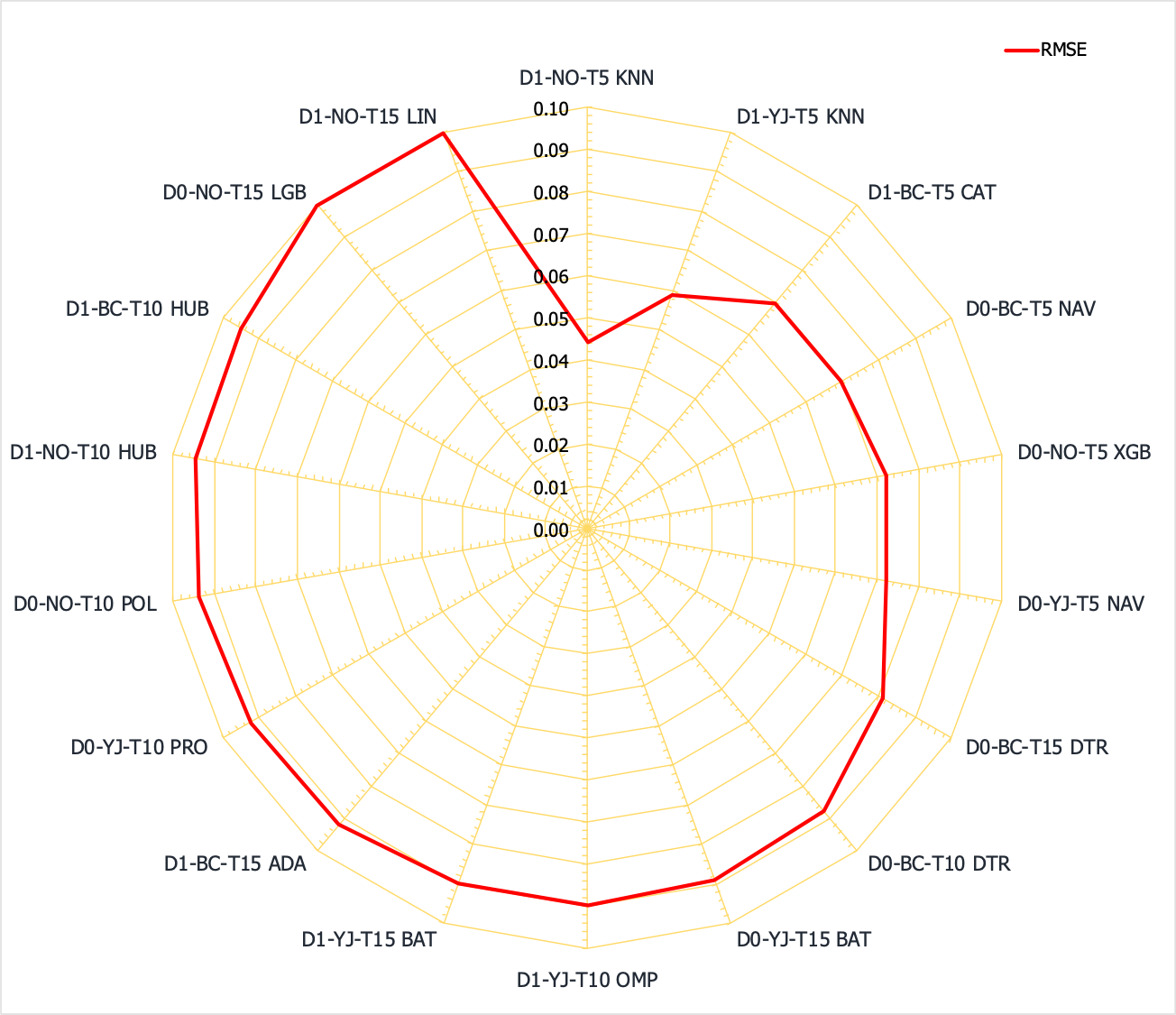}
  \caption{Best Model for Each Run}
  \label{fig:best-model-each-run-T=5-15}
\end{center}
\end{figure}

As stated earlier, six combinations of the data preparation varieties form six different experimental set-ups. For each of these experimental set-ups, all the models are trained, tuned and tested. The result is shown in Figure \ref{fig:best-model-each-run-T=5-15}. It can only be inferred that the effect of the combination of data preparation, test-train split, and algorithm used produces complex patterns. 

\subsection{Regression}
As discussed earlier, using a lag period ($W$) the time series can be converted into a regression problem. The regression algorithms available in
$PyCaret$ are listed in Table \ref{tab:MODELS} of Appendix.
Since the time series methods have been discussed in detail, only the highlights of the results from the regression algorithms are provided.  

\subsubsection{Effect of Data Preparation}
\begin{figure}[H]
\begin{center}
  \includegraphics[width=1\textwidth]{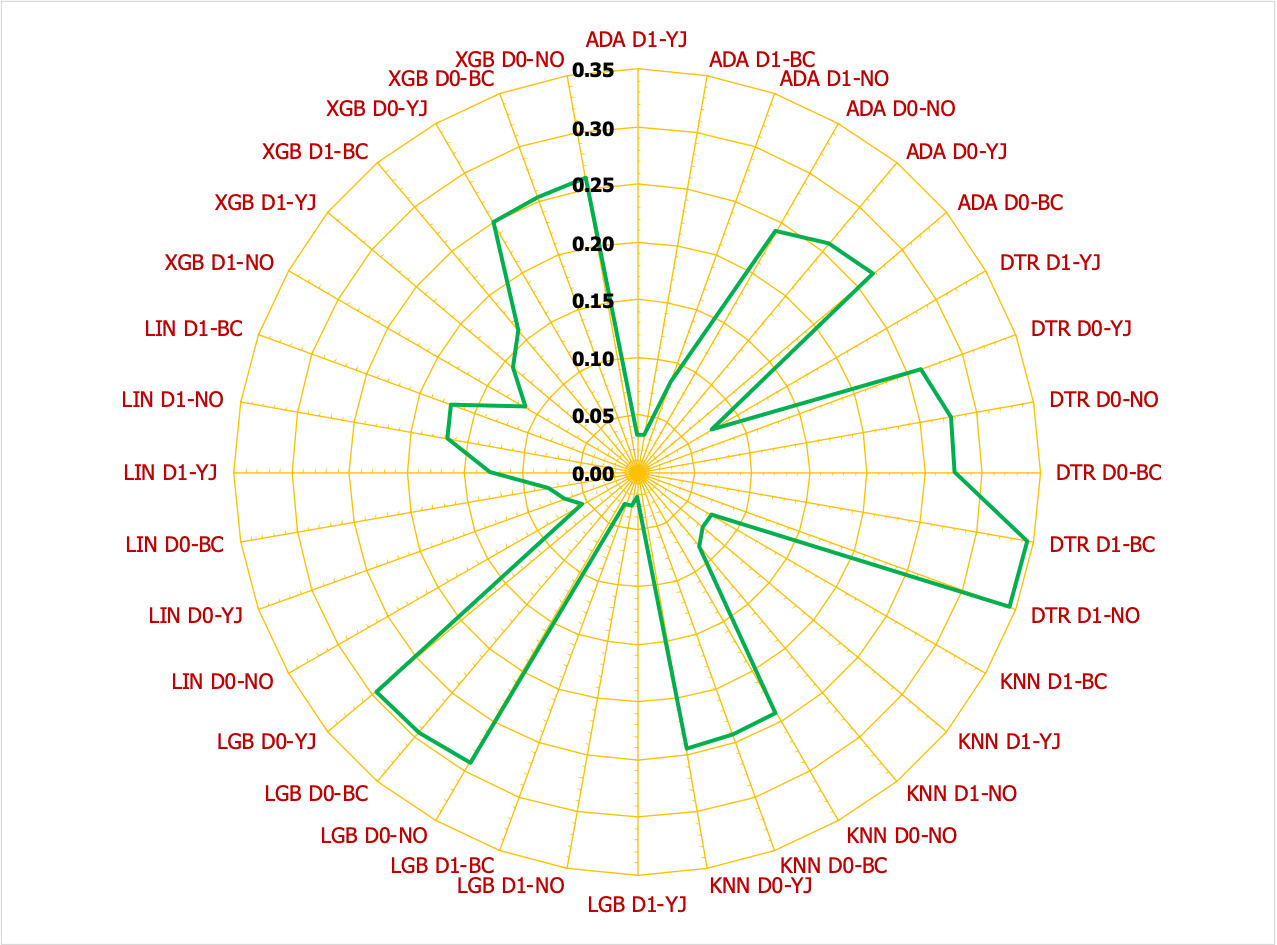}
  \caption{Effects of Data Preparation on the Regression Algorithms}
  \label{fig:Data-prep-reg}
\end{center}
\end{figure}

It will be of interest to see the effects of first-order differencing and power transforms on the Regression algorithms as well. A handful of algorithms, out of a total 25 has been selected to demonstrate it. Figure \ref{fig:Data-prep-reg} shows the utility of data preparation.
It is seen that, first-order differencing which de-trends the data is beneficial to a large extent for most of the algorithms, barring DTR and LIN. Also, the combination D1-YJ was found to increase the accuracy dramatically for ADA, DTR, KNN, LGB, but the effect is not so pronounced for XGB.

\begin{figure}[H]
\begin{center}
  \includegraphics[width=1\textwidth]{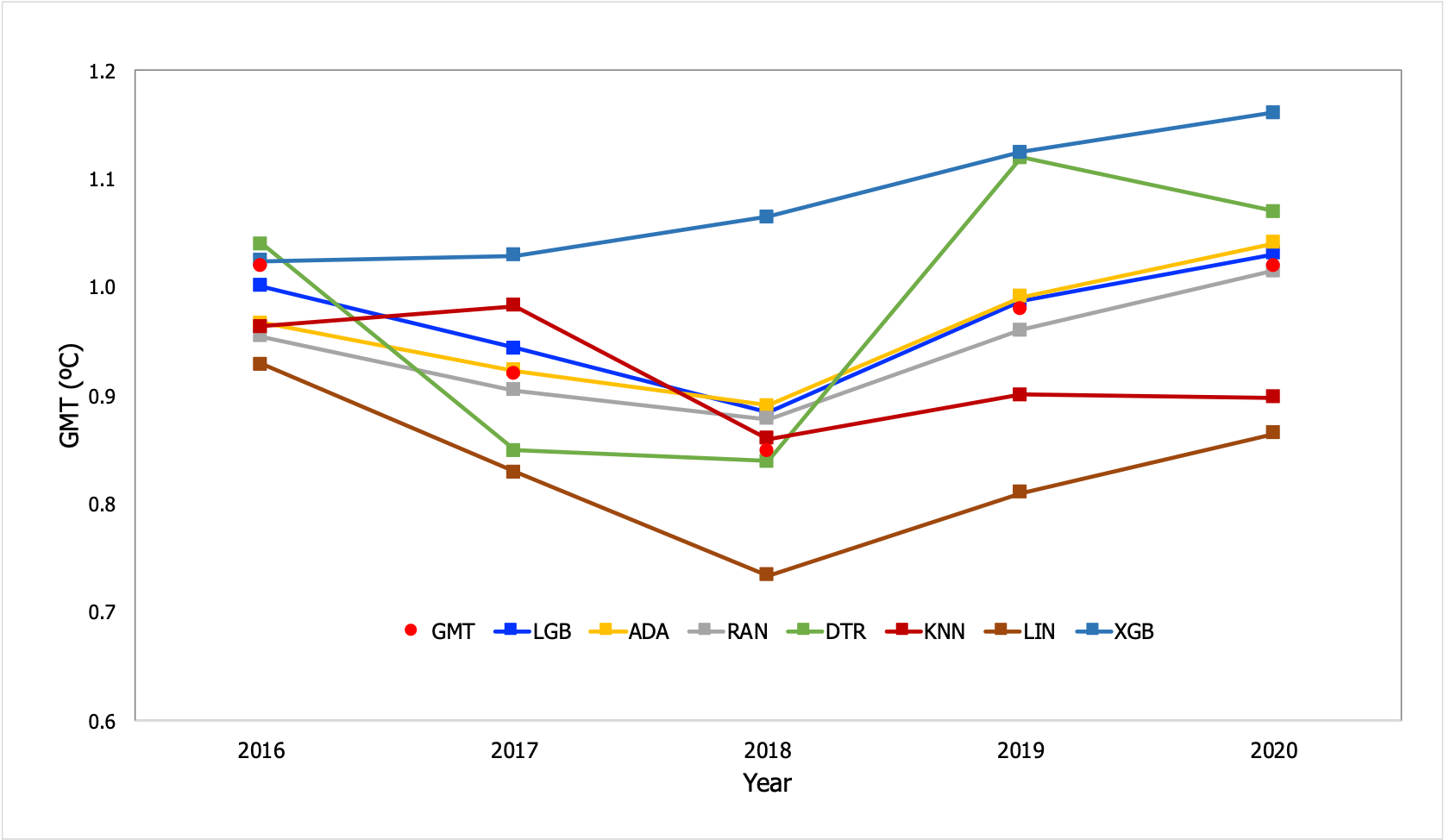}
  \caption{Observed vs Prediction for Regression Algorithms (T = 5)}
  \label{fig:REG-Results}
\end{center}
\end{figure}

The observed vs predicted GMT for a number of regression algorithms for T = 5 are presented in Figure \ref{fig:REG-Results}. The lowest RMSE of 0.02 was observed for {\it{LGB}}.

The Tables \ref{tab:REG-RMSE} and \ref{tab:REG-RMSE of Mean} summarise the top RMSE and the top RMSE (of Mean) values obtained for regression algorithms, respectively. Here the key W-29-D1-YJ-T5 represents a window size of 29, first-order differencing, Yeo-Johnson power transform and the test size of 5. It is seen that LGB scored the lowest RMSE and lowest RMSE (of Mean) for test size equal to five.

\begin{table}[htbp]
  \centering
  \caption{List of Top RMSE Values for Regression Algorithms for Different Data Preparations and Test Sizes}
    \begin{tabular}{|l|l|r|}
    \hline
    \rowcolor[rgb]{ .125,  .216,  .392} \textcolor[rgb]{ 1,  1,  1}{Data Key and Test Size} & \textcolor[rgb]{ 1,  1,  1}{Model Code} & \multicolumn{1}{l|}{\textcolor[rgb]{ 1,  1,  1}{RMSE}} \\
    \hline
    \rowcolor[rgb]{ 1,  .949,  .8} W29-D1-YJ-T5 & LGB   & \cellcolor[rgb]{ 1,  .78,  .808}\textcolor[rgb]{ .612,  0,  .024}{0.0218} \\
    \hline
    \rowcolor[rgb]{ 1,  .949,  .8} W29-D1-NO-T5 & LGB   & 0.0290 \\
    \hline
    \rowcolor[rgb]{ 1,  .949,  .8} W29-D1-BC-T5 & LGB   & 0.0302 \\
    \hline
    \rowcolor[rgb]{ 1,  .949,  .8} W35-D0-YJ-T10 & KRI   & 0.0642 \\
    \hline
    \rowcolor[rgb]{ 1,  .949,  .8} W38-D0-BC-T10 & OMP   & 0.0716 \\
    \hline
    \rowcolor[rgb]{ 1,  .949,  .8} W35-D0-NO-T10 & KRI   & 0.0718 \\
    \hline
    \rowcolor[rgb]{ 1,  .949,  .8} W32-D0-YJ-T15 & BAR   & 0.0729 \\
    \hline
    \rowcolor[rgb]{ 1,  .949,  .8} W39-D1-NO-T15 & XGB   & 0.0737 \\
    \hline
    \rowcolor[rgb]{ 1,  .949,  .8} W41-D1-BC-T15 & XGB   & 0.0744 \\
   \hline
    \end{tabular}%
  \label{tab:REG-RMSE}%
\end{table}%

\begin{table}[htbp]
  \centering
  \caption{List of Top RMSE (of Mean) Values for Regression Algorithms for Different Data Preparations and Test Sizes}
    \begin{tabular}{|l|l|r|}
    \hline
    \rowcolor[rgb]{ .125,  .216,  .392} \textcolor[rgb]{ 1,  1,  1}{Data Key and Test Size} & \textcolor[rgb]{ 1,  1,  1}{Model Code} & \multicolumn{1}{l|}{\textcolor[rgb]{ 1,  1,  1}{RMSE (of Mean)}} \\
    \hline
    \rowcolor[rgb]{ 1,  .949,  .8} W32-D1-YJ-T5 & LGB   & \cellcolor[rgb]{ 1,  .78,  .808}\textcolor[rgb]{ .612,  0,  .024}{0.00002} \\
    \hline
    \rowcolor[rgb]{ 1,  .949,  .8} W45-D1-NO-T5 & MLP   & 0.00009 \\
    \hline
    \rowcolor[rgb]{ 1,  .949,  .8} W42-D0-BC-T5 & ARD   & 0.00050 \\
    \hline
    \rowcolor[rgb]{ 1,  .949,  .8} W35-D0-YJ-T10 & KRI   & 0.01389 \\
    \hline
    \rowcolor[rgb]{ 1,  .949,  .8} W38-D0-BC-T10 & OMP   & 0.00284 \\
    \hline
    \rowcolor[rgb]{ 1,  .949,  .8} W35-D0-NO-T10 & KRI   & 0.02009 \\
    \hline
    \rowcolor[rgb]{ 1,  .949,  .8} W44-D1-NO-T15 & ELN   & 0.00017 \\
    \hline
    \rowcolor[rgb]{ 1,  .949,  .8} W30-D0-YJ-T15 & ARD   & 0.00035 \\
    \hline
    \rowcolor[rgb]{ 1,  .949,  .8} W26-D0-BC-T15 & TSR   & 0.00037 \\
   \hline
    \end{tabular}%
  \label{tab:REG-RMSE of Mean}%
\end{table}%

It is observed that employing regression ML techniques significantly improved the results from those obtained by using only the time series methods.
Thus, considering both time series and regression ML methods, the lowest RMSE and the lowest RMSE (of Mean) becomes 0.02 and 0.00002 respectively, which are better results to the knowledge of the authors. The results are good and  certainly will form benchmarks for the more complex ML methods to meet.

\section{Conclusion}
Accurate prediction of GMT or its mean variability over a span of years is a challenge. Annual GMT data during the period 1880-2020 were analysed by treating them both as a time series and also converting it to a regression problem with a systematic implementation of simple ML methods. 

Firstly, the often neglected step of data transformation was implemented. This consisted of differencing, power transforms, and scaling. It was
found that first-order differencing made the data stationary and had a positive impact on the results obtained for the time series and regression ML methods. The highlight of this work is to show that often unused data trasformation can imorove the model's skill substancially. Secondly, the ML algorithms are selected in an unbiased way,  emphasising variety. It was found that for time series methods, simple algorithms like KNN, HUB and LIN did perform better than the more popular algorithms like XGB, LGB and Prophet, where that is not the case with the regression formulation. This points to the value of an unbiased selection of a wider set of algorithms. Thirdly, the benefits of employing a Naïve model have been demonstrated.

It was seen that the performances of the models vary with both the data preparation strategies and test sizes. It is difficult to explain why one model does well compared to all the others. Performance is mostly the result of complex chemistry between three factors: (i) mode of data preparation, (ii) forecast horizon and (iii) suitability of 1 and 2 for the ML algorithm. While it is true that a fair amount of insight can be
obtained for selecting appropriate algorithms from the analysis of data under study, and the knowledge of the algorithm, choosing the right ML algorithm is an iterative process and the success of this depends on the experience of the modellers.  

Some important features of annual GMT data could be observed. Though simple ML methods  may not be  adequate to make a yearly prediction
of GMT, they are sufficient to accurately capture the 5-year mean variability. 

In future work, as mentioned earlier, the GMT time series will be converted to a multivariate regression problem with additional features generated from the GMT's lagged values or by adding exogenous variables. It is to be investigated whether the regressors used are sufficient to accurately predict GMT. It is planned to use all the successful techniques to forecast GMT into the future.

\clearpage

\section{Appendix}
\subsection{Details of Power Transforms}Detailed formulae of Box-Cox and Yeo-Johnson transforms are presented here.
(a) Box Cox \cite{Asar2014} \\

The one-parameter Box-Cox transformation is defined as \\

$$ y_i^{(\lambda)} =
\begin{cases}
 \dfrac{y_i^\lambda - 1}{\lambda} & \text{if } \lambda \neq 0, \\
 \ln y_i & \text{if } \lambda = 0,
\end{cases}
$$

It is to be noted, that Box-Cox transformation cannot be applied to a data series that is not strictly $> 0$. Therefore, the data must be offset by the minimum value in the series and adding a very small number to make all the data greater than zero. \\ 

$$ y^T(t) = y(t) + abs(y_{Min}) + \epsilon $$

where $y_{Min}$ is the minimum value of the entire series and $\epsilon$ is a small positive number. \\ 

(b) Yeo-Johnson \cite{yj2000} \\

The Yeo-Johnson transformation allows for zero and negative values of $y$. \\

$\lambda$ can be any real number, where $\lambda = 1$ produces the identity transformation. The transformation law reads \\  

$$y_i^{(\lambda)} = \begin{cases} ((y_i+1)^\lambda-1)/\lambda                      &  \text{if }\lambda \neq 0, y \geq 0 \\ 
                                \log(y_i + 1)                                    &  \text{if }\lambda =    0, y \geq 0 \\ 
                                -((-y_i + 1)^{(2-\lambda)} - 1) / (2 - \lambda)  &  \text{if }\lambda \neq 2, y <    0 \\ 
                                -\log(-y_i + 1)                                  &  \text{if }\lambda =    2, y <    0 
                  \end{cases} $$

\subsection{Train Cross Validation Split}
\begin{figure}[H]
\begin{center}
  \includegraphics[width=1\textwidth]{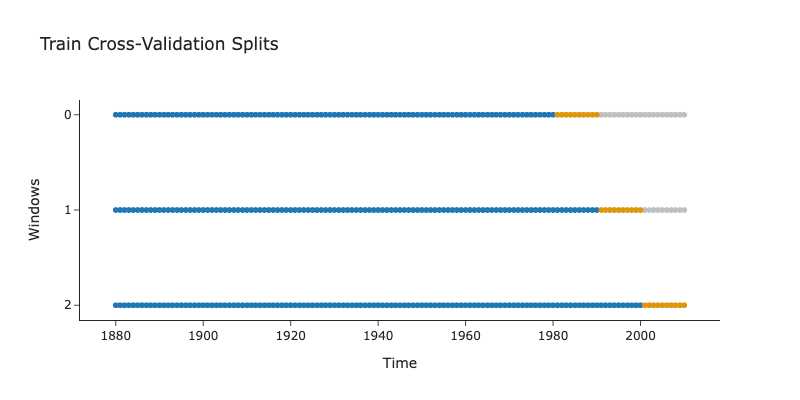}
  \caption{Expanding Training Window with 3 Fold Cross Validation}
  \label{fig:cv}
\end{center}
\end{figure}
\FloatBarrier

\subsection{Predictive Models Used}

The following models listed in Table \ref{tab:MODELS} are considered in this study.  The Time series and Regression algorithms are listed under separate heads.

\begin{table}[htbp]
  \centering
  \caption{Time Series and Regression Models}
    \begin{tabular}{|c|l|c|r|}
    \hline
    \rowcolor[rgb]{ 1,  .949,  .8} \multicolumn{2}{|c|}{Time Series} & \multicolumn{2}{c|}{\cellcolor[rgb]{ .886,  .937,  .855}Regression} \\
   \hline
    \rowcolor[rgb]{ .741,  .843,  .933} \textbf{Code} & \textbf{Name} & \textbf{Code} & \multicolumn{1}{l|}{\textbf{Name}} \\
   \hline
    AAR   & Auto ARIMA & ADA   & \multicolumn{1}{l|}{AdaBoost Regressor} \\
   \hline
    ADA   & AdaBoost & ARD   & \multicolumn{1}{l|}{Automatic Relevance Determination} \\
   \hline
    ARI   & ARIMA & BAR   & \multicolumn{1}{l|}{Bayesian Ridge} \\
   \hline
    BAR   & Bayesian Ridge & CAT   & \multicolumn{1}{l|}{CatBoost Regressor} \\
   \hline
    BAT   & BATS  & DTR   & \multicolumn{1}{l|}{Decision Tree Regressor} \\
   \hline
    CAT   & CatBoost & DUM   & \multicolumn{1}{l|}{Dummy Regressor} \\
   \hline
    CRS   & Croston & ELN   & \multicolumn{1}{l|}{Elastic Net} \\
   \hline
    DTR   & Decision Tree & EXT   & \multicolumn{1}{l|}{Extra Trees Regressor} \\
   \hline
    ELN   & Elastic Net & GRB   & \multicolumn{1}{l|}{Gradient Boosting Regressor} \\
   \hline
    ETS   & ETS   & HUB   & \multicolumn{1}{l|}{Huber Regressor} \\
   \hline
    EXS   & Exponential Smoothing & KNN   & \multicolumn{1}{l|}{K Neighbors Regressor} \\
   \hline
    EXT   & Extra Trees & KRI   & \multicolumn{1}{l|}{Kernel Ridge} \\
   \hline
    GBO   & Gradient Boosting & LAR   & \multicolumn{1}{l|}{Least Angle Regression} \\
   \hline
    GRM   & Grand Means & LAS   & \multicolumn{1}{l|}{Lasso Regression} \\
   \hline
    HUB   & Huber & LGB   & \multicolumn{1}{l|}{Light Gradient Boosting Machine} \\
   \hline
    KNN   & K Neighbors & LIN   & \multicolumn{1}{l|}{Linear Regression} \\
   \hline
    LEA   & Least Angular & LLA   & \multicolumn{1}{l|}{Lasso Least Angle Regression} \\
   \hline
    LGB   & Light Gradient Boosting & MLP   & \multicolumn{1}{l|}{MLP Regressor} \\
   \hline
    LIN   & Linear & OMP   & \multicolumn{1}{l|}{Orthogonal Matching Pursuit} \\
   \hline
    LLA   & Lasso Least Angular & PAR   & \multicolumn{1}{l|}{Passive Aggressive Regressor} \\
   \hline
    LSO   & Lasso & RAN   & \multicolumn{1}{l|}{Random Forest Regressor} \\
   \hline
    NAV   & Naïve & RID   & \multicolumn{1}{l|}{Ridge Regression} \\
   \hline
    OMP   & Orthogonal Matching Pursuit & RSC   & \multicolumn{1}{l|}{Random Sample Consensus} \\
   \hline
    PAG   & Passive Aggressive & SVM   & \multicolumn{1}{l|}{Support Vector Regression} \\
   \hline
    POT   & Polynomial Trend & TSR   & \multicolumn{1}{l|}{TheilSen Regressor} \\
   \hline
    PRO   & Prophet & XGB   & \multicolumn{1}{l|}{Extreme Gradient Boosting} \\
   \hline
    RAN   & Random Forest &       &  \\
   \hline
    RID   & Ridge &       &  \\
   \hline
    TBA   & TBATS &       &  \\
   \hline
    THE   & Theta &       &  \\
   \hline
    XGB   & Extreme Gradient Boosting &       &  \\
    \hline
    \end{tabular}%
  \label{tab:MODELS}%
\end{table}%

\FloatBarrier

\subsection{RMSE Values for All the Models Where No Differencing is Done (D0)}
\begin{table}[!htb]
  \centering
  \caption{Results for No Differencing}
\begin{adjustbox}{width=0.9\columnwidth,center}
    \begin{tabular}{|p{4.085em}|r|r|r|r|r|r|r|r|r|}
   \hline
    \rowcolor[rgb]{ .973,  .796,  .678} \multicolumn{1}{|l|}{Case: D0} & \multicolumn{3}{c|}{\cellcolor[rgb]{ 1,  .902,  .6}D0-BC} & \multicolumn{3}{c|}{\cellcolor[rgb]{ 1,  .902,  .6}D0-NO} & \multicolumn{3}{c|}{\cellcolor[rgb]{ 1,  .902,  .6}D0-YJ} \\
   \hline
    \rowcolor[rgb]{ .125,  .216,  .392} \multicolumn{1}{|l|}{\textcolor[rgb]{ 1,  1,  1}{Model Code}} & \multicolumn{1}{c|}{\cellcolor[rgb]{ .851,  .851,  .851}5} & \multicolumn{1}{c|}{\cellcolor[rgb]{ .851,  .882,  .949}10} & \multicolumn{1}{c|}{\cellcolor[rgb]{ .988,  .894,  .839}15} & \multicolumn{1}{c|}{\cellcolor[rgb]{ .851,  .851,  .851}5} & \multicolumn{1}{c|}{\cellcolor[rgb]{ .851,  .882,  .949}10} & \multicolumn{1}{c|}{\cellcolor[rgb]{ .988,  .894,  .839}15} & \multicolumn{1}{c|}{\cellcolor[rgb]{ .851,  .851,  .851}5} & \multicolumn{1}{c|}{\cellcolor[rgb]{ .851,  .882,  .949}10} & \multicolumn{1}{c|}{\cellcolor[rgb]{ .988,  .894,  .839}15} \\
    \hline
    \rowcolor[rgb]{ .886,  .937,  .855} AAR   & \cellcolor[rgb]{ 1,  1,  1}0.162 & \cellcolor[rgb]{ 1,  1,  1}0.181 & \cellcolor[rgb]{ 1,  1,  1}0.243 & \cellcolor[rgb]{ 1,  1,  1}0.165 & \cellcolor[rgb]{ 1,  1,  1}0.191 & \cellcolor[rgb]{ 1,  1,  1}0.179 & \cellcolor[rgb]{ 1,  1,  1}0.148 & \cellcolor[rgb]{ 1,  1,  1}0.159 & \cellcolor[rgb]{ 1,  1,  1}0.248 \\
   \hline
    \rowcolor[rgb]{ .886,  .937,  .855} ADA   & \cellcolor[rgb]{ 1,  1,  1}0.112 & \cellcolor[rgb]{ 1,  1,  1}0.090 & \cellcolor[rgb]{ 1,  1,  1}0.321 & \cellcolor[rgb]{ 1,  1,  1}0.129 & \cellcolor[rgb]{ 1,  1,  1}0.148 & \cellcolor[rgb]{ 1,  1,  1}0.423 & \cellcolor[rgb]{ 1,  1,  1}0.073 & \cellcolor[rgb]{ 1,  1,  1}0.121 & \cellcolor[rgb]{ 1,  1,  1}0.143 \\
   \hline
    \rowcolor[rgb]{ .886,  .937,  .855} ARI   & \cellcolor[rgb]{ 1,  1,  1}0.173 & \cellcolor[rgb]{ 1,  1,  1}0.156 & \cellcolor[rgb]{ 1,  1,  1}0.125 & \cellcolor[rgb]{ 1,  1,  1}0.140 & \cellcolor[rgb]{ 1,  1,  1}0.213 & \cellcolor[rgb]{ 1,  1,  1}0.139 & \cellcolor[rgb]{ 1,  1,  1}0.155 & \cellcolor[rgb]{ 1,  1,  1}0.147 & \cellcolor[rgb]{ 1,  1,  1}0.113 \\
   \hline
    \rowcolor[rgb]{ .886,  .937,  .855} BAR   & \cellcolor[rgb]{ 1,  1,  1}0.075 & \cellcolor[rgb]{ 1,  1,  1}0.093 & \cellcolor[rgb]{ 1,  1,  1}0.336 & \cellcolor[rgb]{ 1,  1,  1}0.146 & \cellcolor[rgb]{ 1,  1,  1}0.151 & \cellcolor[rgb]{ 1,  1,  1}0.140 & \cellcolor[rgb]{ 1,  1,  1}0.095 & \cellcolor[rgb]{ 1,  1,  1}0.147 & \cellcolor[rgb]{ 1,  1,  1}0.329 \\
   \hline
    \rowcolor[rgb]{ .886,  .937,  .855} BAT   & \cellcolor[rgb]{ 1,  1,  1}0.219 & \cellcolor[rgb]{ 1,  1,  1}0.208 & \cellcolor[rgb]{ 1,  1,  1}0.197 & \cellcolor[rgb]{ 1,  1,  1}0.130 & \cellcolor[rgb]{ 1,  1,  1}0.155 & \cellcolor[rgb]{ 1,  1,  1}0.214 & \cellcolor[rgb]{ 1,  1,  1}0.111 & \cellcolor[rgb]{ 1,  1,  1}0.113 & \cellcolor[rgb]{ 1,  .949,  .8}\textcolor[rgb]{ 1,  0,  0}{\textbf{0.089}} \\
   \hline
    \rowcolor[rgb]{ .886,  .937,  .855} CAT   & \cellcolor[rgb]{ 1,  1,  1}0.092 & \cellcolor[rgb]{ 1,  1,  1}0.091 & \cellcolor[rgb]{ 1,  1,  1}0.161 & \cellcolor[rgb]{ 1,  1,  1}0.162 & \cellcolor[rgb]{ 1,  1,  1}0.140 & \cellcolor[rgb]{ 1,  1,  1}0.258 & \cellcolor[rgb]{ 1,  1,  1}0.087 & \cellcolor[rgb]{ 1,  1,  1}0.134 & \cellcolor[rgb]{ 1,  1,  1}0.141 \\
   \hline
    \rowcolor[rgb]{ .886,  .937,  .855} CRS   & \cellcolor[rgb]{ 1,  1,  1}0.168 & \cellcolor[rgb]{ 1,  1,  1}0.206 & \cellcolor[rgb]{ 1,  1,  1}0.208 & \cellcolor[rgb]{ 1,  1,  1}0.143 & \cellcolor[rgb]{ 1,  1,  1}0.200 & \cellcolor[rgb]{ 1,  1,  1}0.182 & \cellcolor[rgb]{ 1,  1,  1}0.170 & \cellcolor[rgb]{ 1,  1,  1}0.207 & \cellcolor[rgb]{ 1,  1,  1}0.209 \\
   \hline
    \rowcolor[rgb]{ .886,  .937,  .855} DTR   & \cellcolor[rgb]{ 1,  1,  1}0.090 & \cellcolor[rgb]{ 1,  .949,  .8}\textcolor[rgb]{ 1,  0,  0}{\textbf{0.088}} & \cellcolor[rgb]{ 1,  .949,  .8}\textcolor[rgb]{ 1,  0,  0}{\textbf{0.081}} & \cellcolor[rgb]{ 1,  1,  1}0.108 & \cellcolor[rgb]{ 1,  1,  1}0.144 & \cellcolor[rgb]{ 1,  1,  1}0.428 & \cellcolor[rgb]{ 1,  1,  1}0.087 & \cellcolor[rgb]{ 1,  1,  1}0.112 & \cellcolor[rgb]{ 1,  1,  1}0.135 \\
   \hline
    \rowcolor[rgb]{ .886,  .937,  .855} ELN   & \cellcolor[rgb]{ 1,  1,  1}0.091 & \cellcolor[rgb]{ 1,  1,  1}0.091 & \cellcolor[rgb]{ 1,  1,  1}0.351 & \cellcolor[rgb]{ 1,  1,  1}0.178 & \cellcolor[rgb]{ 1,  1,  1}0.149 & \cellcolor[rgb]{ 1,  1,  1}0.140 & \cellcolor[rgb]{ 1,  1,  1}0.093 & \cellcolor[rgb]{ 1,  1,  1}0.146 & \cellcolor[rgb]{ 1,  1,  1}0.346 \\
   \hline
    \rowcolor[rgb]{ .886,  .937,  .855} ETS   & \cellcolor[rgb]{ 1,  1,  1}0.136 & \cellcolor[rgb]{ 1,  1,  1}0.167 & \cellcolor[rgb]{ 1,  1,  1}0.147 & \cellcolor[rgb]{ 1,  1,  1}0.143 & \cellcolor[rgb]{ 1,  1,  1}0.181 & \cellcolor[rgb]{ 1,  1,  1}0.164 & \cellcolor[rgb]{ 1,  1,  1}0.125 & \cellcolor[rgb]{ 1,  1,  1}0.153 & \cellcolor[rgb]{ 1,  1,  1}0.130 \\
   \hline
    \rowcolor[rgb]{ .886,  .937,  .855} EXP   & \cellcolor[rgb]{ 1,  1,  1}0.136 & \cellcolor[rgb]{ 1,  1,  1}0.167 & \cellcolor[rgb]{ 1,  1,  1}0.147 & \cellcolor[rgb]{ 1,  1,  1}0.143 & \cellcolor[rgb]{ 1,  1,  1}0.181 & \cellcolor[rgb]{ 1,  1,  1}0.164 & \cellcolor[rgb]{ 1,  1,  1}0.125 & \cellcolor[rgb]{ 1,  1,  1}0.153 & \cellcolor[rgb]{ 1,  1,  1}0.129 \\
   \hline
    \rowcolor[rgb]{ .886,  .937,  .855} EXT   & \cellcolor[rgb]{ 1,  1,  1}0.090 & \cellcolor[rgb]{ 1,  1,  1}0.096 & \cellcolor[rgb]{ 1,  1,  1}0.094 & \cellcolor[rgb]{ 1,  1,  1}0.161 & \cellcolor[rgb]{ 1,  1,  1}0.141 & \cellcolor[rgb]{ 1,  1,  1}0.135 & \cellcolor[rgb]{ 1,  1,  1}0.095 & \cellcolor[rgb]{ 1,  1,  1}0.125 & \cellcolor[rgb]{ 1,  1,  1}0.135 \\
   \hline
    \rowcolor[rgb]{ .886,  .937,  .855} GRB   & \cellcolor[rgb]{ 1,  1,  1}0.092 & \cellcolor[rgb]{ 1,  1,  1}0.091 & \cellcolor[rgb]{ 1,  1,  1}0.360 & \cellcolor[rgb]{ 1,  1,  1}0.135 & \cellcolor[rgb]{ 1,  1,  1}0.182 & \cellcolor[rgb]{ 1,  1,  1}0.158 & \cellcolor[rgb]{ 1,  1,  1}0.093 & \cellcolor[rgb]{ 1,  1,  1}0.136 & \cellcolor[rgb]{ 1,  1,  1}0.385 \\
   \hline
    \rowcolor[rgb]{ .886,  .937,  .855} HUB   & \cellcolor[rgb]{ 1,  1,  1}0.070 & \cellcolor[rgb]{ 1,  1,  1}0.091 & \cellcolor[rgb]{ 1,  1,  1}0.303 & \cellcolor[rgb]{ 1,  1,  1}0.142 & \cellcolor[rgb]{ 1,  1,  1}0.147 & \cellcolor[rgb]{ 1,  1,  1}0.137 & \cellcolor[rgb]{ 1,  1,  1}0.089 & \cellcolor[rgb]{ 1,  1,  1}0.135 & \cellcolor[rgb]{ 1,  1,  1}0.298 \\
   \hline
    \rowcolor[rgb]{ .886,  .937,  .855} KNN   & \cellcolor[rgb]{ 1,  1,  1}0.220 & \cellcolor[rgb]{ 1,  1,  1}0.227 & \cellcolor[rgb]{ 1,  1,  1}0.225 & \cellcolor[rgb]{ 1,  1,  1}0.215 & \cellcolor[rgb]{ 1,  1,  1}0.203 & \cellcolor[rgb]{ 1,  1,  1}0.440 & \cellcolor[rgb]{ 1,  1,  1}0.189 & \cellcolor[rgb]{ 1,  1,  1}0.141 & \cellcolor[rgb]{ 1,  1,  1}0.152 \\
   \hline
    \rowcolor[rgb]{ .886,  .937,  .855} LAR   & \cellcolor[rgb]{ 1,  1,  1}0.077 & \cellcolor[rgb]{ 1,  1,  1}0.093 & \cellcolor[rgb]{ 1,  1,  1}0.334 & \cellcolor[rgb]{ 1,  1,  1}0.145 & \cellcolor[rgb]{ 1,  1,  1}0.151 & \cellcolor[rgb]{ 1,  1,  1}0.140 & \cellcolor[rgb]{ 1,  1,  1}0.091 & \cellcolor[rgb]{ 1,  1,  1}0.140 & \cellcolor[rgb]{ 1,  1,  1}0.329 \\
   \hline
    \rowcolor[rgb]{ .886,  .937,  .855} LAS   & \cellcolor[rgb]{ 1,  1,  1}0.092 & \cellcolor[rgb]{ 1,  1,  1}0.091 & \cellcolor[rgb]{ 1,  1,  1}0.354 & \cellcolor[rgb]{ 1,  1,  1}0.150 & \cellcolor[rgb]{ 1,  1,  1}0.149 & \cellcolor[rgb]{ 1,  1,  1}0.144 & \cellcolor[rgb]{ 1,  1,  1}0.093 & \cellcolor[rgb]{ 1,  1,  1}0.148 & \cellcolor[rgb]{ 1,  1,  1}0.349 \\
   \hline
    \rowcolor[rgb]{ .886,  .937,  .855} LGB   & \cellcolor[rgb]{ 1,  1,  1}0.092 & \cellcolor[rgb]{ 1,  1,  1}0.091 & \cellcolor[rgb]{ 1,  1,  1}0.131 & \cellcolor[rgb]{ 1,  1,  1}0.165 & \cellcolor[rgb]{ 1,  1,  1}0.150 & \cellcolor[rgb]{ 1,  .949,  .8}\textcolor[rgb]{ 1,  0,  0}{\textbf{0.100}} & \cellcolor[rgb]{ 1,  1,  1}0.100 & \cellcolor[rgb]{ 1,  1,  1}0.132 & \cellcolor[rgb]{ 1,  1,  1}0.104 \\
   \hline
    \rowcolor[rgb]{ .886,  .937,  .855} LIN   & \cellcolor[rgb]{ 1,  1,  1}0.075 & \cellcolor[rgb]{ 1,  1,  1}0.093 & \cellcolor[rgb]{ 1,  1,  1}0.334 & \cellcolor[rgb]{ 1,  1,  1}0.145 & \cellcolor[rgb]{ 1,  1,  1}0.151 & \cellcolor[rgb]{ 1,  1,  1}0.140 & \cellcolor[rgb]{ 1,  1,  1}0.095 & \cellcolor[rgb]{ 1,  1,  1}0.147 & \cellcolor[rgb]{ 1,  1,  1}0.327 \\
   \hline
    \rowcolor[rgb]{ .886,  .937,  .855} LLA   & \cellcolor[rgb]{ 1,  1,  1}0.077 & \cellcolor[rgb]{ 1,  1,  1}0.093 & \cellcolor[rgb]{ 1,  1,  1}0.334 & \cellcolor[rgb]{ 1,  1,  1}0.146 & \cellcolor[rgb]{ 1,  1,  1}0.151 & \cellcolor[rgb]{ 1,  1,  1}0.140 & \cellcolor[rgb]{ 1,  1,  1}0.091 & \cellcolor[rgb]{ 1,  1,  1}0.140 & \cellcolor[rgb]{ 1,  1,  1}0.329 \\
    \hline
    \rowcolor[rgb]{ .663,  .816,  .557} NAV   & \cellcolor[rgb]{ 1,  .949,  .8}\textcolor[rgb]{ 1,  0,  0}{\textbf{0.070}} & \cellcolor[rgb]{ 1,  1,  1}0.141 & \cellcolor[rgb]{ 1,  1,  1}0.123 & \cellcolor[rgb]{ 1,  1,  1}0.073 & \cellcolor[rgb]{ 1,  1,  1}0.154 & \cellcolor[rgb]{ 1,  1,  1}0.137 & \cellcolor[rgb]{ 1,  0,  0}\textcolor[rgb]{ 1,  1,  1}{\textbf{0.067}} & \cellcolor[rgb]{ 1,  1,  1}0.129 & \cellcolor[rgb]{ 1,  1,  1}0.111 \\
    \hline
    \rowcolor[rgb]{ .886,  .937,  .855} OMP   & \cellcolor[rgb]{ 1,  1,  1}0.293 & \cellcolor[rgb]{ 1,  1,  1}0.340 & \cellcolor[rgb]{ 1,  1,  1}0.340 & \cellcolor[rgb]{ 1,  1,  1}0.264 & \cellcolor[rgb]{ 1,  1,  1}0.363 & \cellcolor[rgb]{ 1,  1,  1}0.377 & \cellcolor[rgb]{ 1,  1,  1}0.285 & \cellcolor[rgb]{ 1,  1,  1}0.331 & \cellcolor[rgb]{ 1,  1,  1}0.333 \\
   \hline
    \rowcolor[rgb]{ .886,  .937,  .855} POL   & \cellcolor[rgb]{ 1,  1,  1}0.089 & \cellcolor[rgb]{ 1,  1,  1}0.090 & \cellcolor[rgb]{ 1,  1,  1}0.354 & \cellcolor[rgb]{ 1,  1,  1}0.100 & \cellcolor[rgb]{ 1,  .949,  .8}\textcolor[rgb]{ 1,  0,  0}{\textbf{0.094}} & \cellcolor[rgb]{ 1,  1,  1}0.143 & \cellcolor[rgb]{ 1,  1,  1}0.093 & \cellcolor[rgb]{ 1,  1,  1}0.148 & \cellcolor[rgb]{ 1,  1,  1}0.348 \\
    \hline
    \rowcolor[rgb]{ .886,  .937,  .855} PRO   & \cellcolor[rgb]{ 1,  1,  1}0.134 & \cellcolor[rgb]{ 1,  1,  1}0.107 & \cellcolor[rgb]{ 1,  1,  1}0.272 & \cellcolor[rgb]{ 1,  1,  1}0.159 & \cellcolor[rgb]{ 1,  1,  1}0.119 & \cellcolor[rgb]{ 1,  1,  1}0.103 & \cellcolor[rgb]{ 1,  1,  1}0.111 & \cellcolor[rgb]{ 1,  .949,  .8}\textcolor[rgb]{ 1,  0,  0}{\textbf{0.093}} & \cellcolor[rgb]{ 1,  1,  1}0.222 \\
   \hline
    \rowcolor[rgb]{ .886,  .937,  .855} RAN   & \cellcolor[rgb]{ 1,  1,  1}0.092 & \cellcolor[rgb]{ 1,  1,  1}0.091 & \cellcolor[rgb]{ 1,  1,  1}0.085 & \cellcolor[rgb]{ 1,  1,  1}0.151 & \cellcolor[rgb]{ 1,  1,  1}0.149 & \cellcolor[rgb]{ 1,  1,  1}0.144 & \cellcolor[rgb]{ 1,  1,  1}0.093 & \cellcolor[rgb]{ 1,  1,  1}0.148 & \cellcolor[rgb]{ 1,  1,  1}0.285 \\
   \hline
    \rowcolor[rgb]{ .886,  .937,  .855} RID   & \cellcolor[rgb]{ 1,  1,  1}0.091 & \cellcolor[rgb]{ 1,  1,  1}0.091 & \cellcolor[rgb]{ 1,  1,  1}0.339 & \cellcolor[rgb]{ 1,  1,  1}0.153 & \cellcolor[rgb]{ 1,  1,  1}0.151 & \cellcolor[rgb]{ 1,  1,  1}0.141 & \cellcolor[rgb]{ 1,  1,  1}0.093 & \cellcolor[rgb]{ 1,  1,  1}0.147 & \cellcolor[rgb]{ 1,  1,  1}0.334 \\
   \hline
    \rowcolor[rgb]{ .886,  .937,  .855} TBA   & \cellcolor[rgb]{ 1,  1,  1}0.219 & \cellcolor[rgb]{ 1,  1,  1}0.208 & \cellcolor[rgb]{ 1,  1,  1}0.197 & \cellcolor[rgb]{ 1,  1,  1}0.130 & \cellcolor[rgb]{ 1,  1,  1}0.155 & \cellcolor[rgb]{ 1,  1,  1}0.214 & \cellcolor[rgb]{ 1,  1,  1}0.111 & \cellcolor[rgb]{ 1,  1,  1}0.113 & \cellcolor[rgb]{ 1,  .949,  .8}\textcolor[rgb]{ 1,  0,  0}{\textbf{0.089}} \\
   \hline
    \rowcolor[rgb]{ .886,  .937,  .855} THE   & \cellcolor[rgb]{ 1,  1,  1}0.139 & \cellcolor[rgb]{ 1,  1,  1}0.181 & \cellcolor[rgb]{ 1,  1,  1}0.168 & \cellcolor[rgb]{ 1,  1,  1}0.145 & \cellcolor[rgb]{ 1,  1,  1}0.196 & \cellcolor[rgb]{ 1,  1,  1}0.186 & \cellcolor[rgb]{ 1,  1,  1}0.129 & \cellcolor[rgb]{ 1,  1,  1}0.170 & \cellcolor[rgb]{ 1,  1,  1}0.155 \\
   \hline
    \rowcolor[rgb]{ .886,  .937,  .855} XGB   & \cellcolor[rgb]{ 1,  1,  1}0.138 & \cellcolor[rgb]{ 1,  1,  1}0.140 & \cellcolor[rgb]{ 1,  1,  1}0.162 & \cellcolor[rgb]{ 1,  .949,  .8}\textcolor[rgb]{ 1,  0,  0}{\textbf{0.072}} & \cellcolor[rgb]{ 1,  1,  1}0.106 & \cellcolor[rgb]{ 1,  1,  1}0.292 & \cellcolor[rgb]{ 1,  1,  1}0.119 & \cellcolor[rgb]{ 1,  1,  1}0.115 & \cellcolor[rgb]{ 1,  1,  1}0.121 \\
    \hline
    \end{tabular}%
  \label{tab:RMSE-D0}%
  \end{adjustbox}
\end{table}%
\FloatBarrier

\subsection{RMSE Values for all the Models where First-Order Differencing is Done (D1)}
\begin{table}[!htb]
  \centering
  \caption{Results for First-Order Differencing}
\begin{adjustbox}{width=0.9\columnwidth,center}
    \begin{tabular}{|p{4.915em}|r|r|r|r|r|r|r|r|r|}
   \hline
    \rowcolor[rgb]{ .973,  .796,  .678} \multicolumn{1}{|l|}{Case: D1} & \multicolumn{3}{c|}{\cellcolor[rgb]{ 1,  .902,  .6}D1-BC} & \multicolumn{3}{c|}{\cellcolor[rgb]{ 1,  .902,  .6}D1-NO} & \multicolumn{3}{c|}{\cellcolor[rgb]{ 1,  .902,  .6}D1-YJ} \\
   \hline
    \rowcolor[rgb]{ .125,  .216,  .392} \multicolumn{1}{|l|}{\textcolor[rgb]{ 1,  1,  1}{Model Code}} & \multicolumn{1}{c|}{\cellcolor[rgb]{ .851,  .851,  .851}5} & \multicolumn{1}{c|}{\cellcolor[rgb]{ .851,  .882,  .949}10} & \multicolumn{1}{c|}{\cellcolor[rgb]{ .988,  .894,  .839}15} & \multicolumn{1}{c|}{\cellcolor[rgb]{ .851,  .851,  .851}5} & \multicolumn{1}{c|}{\cellcolor[rgb]{ .851,  .882,  .949}10} & \multicolumn{1}{c|}{\cellcolor[rgb]{ .988,  .894,  .839}15} & \multicolumn{1}{c|}{\cellcolor[rgb]{ .851,  .851,  .851}5} & \multicolumn{1}{c|}{\cellcolor[rgb]{ .851,  .882,  .949}10} & \multicolumn{1}{c|}{\cellcolor[rgb]{ .988,  .894,  .839}15} \\
    \hline
    \rowcolor[rgb]{ .886,  .937,  .855} AAR   & \cellcolor[rgb]{ 1,  1,  1}0.160 & \cellcolor[rgb]{ 1,  1,  1}0.181 & \cellcolor[rgb]{ 1,  1,  1}0.163 & \cellcolor[rgb]{ 1,  1,  1}0.158 & \cellcolor[rgb]{ 1,  1,  1}0.177 & \cellcolor[rgb]{ 1,  1,  1}0.159 & \cellcolor[rgb]{ 1,  1,  1}0.142 & \cellcolor[rgb]{ 1,  1,  1}0.150 & \cellcolor[rgb]{ 1,  1,  1}0.124 \\
   \hline
    \rowcolor[rgb]{ .886,  .937,  .855} ADA   & \cellcolor[rgb]{ 1,  1,  1}0.072 & \cellcolor[rgb]{ 1,  1,  1}0.167 & \cellcolor[rgb]{ 1,  .949,  .8}\textcolor[rgb]{ 1,  0,  0}{\textbf{0.092}} & \cellcolor[rgb]{ 1,  1,  1}0.071 & \cellcolor[rgb]{ 1,  1,  1}0.110 & \cellcolor[rgb]{ 1,  1,  1}0.130 & \cellcolor[rgb]{ 1,  1,  1}0.077 & \cellcolor[rgb]{ 1,  1,  1}0.140 & \cellcolor[rgb]{ 1,  1,  1}0.131 \\
   \hline
    \rowcolor[rgb]{ .886,  .937,  .855} ARI   & \cellcolor[rgb]{ 1,  1,  1}0.101 & \cellcolor[rgb]{ 1,  1,  1}0.107 & \cellcolor[rgb]{ 1,  1,  1}0.098 & \cellcolor[rgb]{ 1,  1,  1}0.100 & \cellcolor[rgb]{ 1,  1,  1}0.105 & \cellcolor[rgb]{ 1,  1,  1}0.100 & \cellcolor[rgb]{ 1,  1,  1}0.093 & \cellcolor[rgb]{ 1,  1,  1}0.090 & \cellcolor[rgb]{ 1,  1,  1}0.130 \\
   \hline
    \rowcolor[rgb]{ .886,  .937,  .855} BAR   & \cellcolor[rgb]{ 1,  1,  1}0.074 & \cellcolor[rgb]{ 1,  1,  1}0.100 & \cellcolor[rgb]{ 1,  1,  1}0.100 & \cellcolor[rgb]{ 1,  1,  1}0.074 & \cellcolor[rgb]{ 1,  1,  1}0.099 & \cellcolor[rgb]{ 1,  1,  1}0.102 & \cellcolor[rgb]{ 1,  1,  1}0.075 & \cellcolor[rgb]{ 1,  1,  1}0.093 & \cellcolor[rgb]{ 1,  1,  1}0.131 \\
   \hline
    \rowcolor[rgb]{ .886,  .937,  .855} BAT   & \cellcolor[rgb]{ 1,  1,  1}0.077 & \cellcolor[rgb]{ 1,  1,  1}0.122 & \cellcolor[rgb]{ 1,  1,  1}0.106 & \cellcolor[rgb]{ 1,  1,  1}0.077 & \cellcolor[rgb]{ 1,  1,  1}0.135 & \cellcolor[rgb]{ 1,  1,  1}0.106 & \cellcolor[rgb]{ 1,  1,  1}0.106 & \cellcolor[rgb]{ 1,  1,  1}0.102 & \cellcolor[rgb]{ 1,  .949,  .8}\textcolor[rgb]{ 1,  0,  0}{\textbf{0.090}} \\
    \rowcolor[rgb]{ .886,  .937,  .855} CAT   & \cellcolor[rgb]{ 1,  .949,  .8}\textcolor[rgb]{ 1,  0,  0}{\textbf{0.069}} & \cellcolor[rgb]{ 1,  1,  1}0.100 & \cellcolor[rgb]{ 1,  1,  1}0.137 & \cellcolor[rgb]{ 1,  1,  1}0.078 & \cellcolor[rgb]{ 1,  1,  1}0.099 & \cellcolor[rgb]{ 1,  1,  1}0.111 & \cellcolor[rgb]{ 1,  1,  1}0.076 & \cellcolor[rgb]{ 1,  1,  1}0.096 & \cellcolor[rgb]{ 1,  1,  1}0.152 \\
   \hline
    \rowcolor[rgb]{ .886,  .937,  .855} CRS   & \cellcolor[rgb]{ 1,  1,  1}0.154 & \cellcolor[rgb]{ 1,  1,  1}0.218 & \cellcolor[rgb]{ 1,  1,  1}0.420 & \cellcolor[rgb]{ 1,  1,  1}0.155 & \cellcolor[rgb]{ 1,  1,  1}0.220 & \cellcolor[rgb]{ 1,  1,  1}0.423 & \cellcolor[rgb]{ 1,  1,  1}0.164 & \cellcolor[rgb]{ 1,  1,  1}0.238 & \cellcolor[rgb]{ 1,  1,  1}0.451 \\
   \hline
    \rowcolor[rgb]{ .886,  .937,  .855} DTR   & \cellcolor[rgb]{ 1,  1,  1}0.100 & \cellcolor[rgb]{ 1,  1,  1}0.099 & \cellcolor[rgb]{ 1,  1,  1}0.145 & \cellcolor[rgb]{ 1,  1,  1}0.084 & \cellcolor[rgb]{ 1,  1,  1}0.098 & \cellcolor[rgb]{ 1,  1,  1}0.149 & \cellcolor[rgb]{ 1,  1,  1}0.140 & \cellcolor[rgb]{ 1,  1,  1}0.096 & \cellcolor[rgb]{ 1,  1,  1}0.123 \\
   \hline
    \rowcolor[rgb]{ .886,  .937,  .855} ELN   & \cellcolor[rgb]{ 1,  1,  1}0.072 & \cellcolor[rgb]{ 1,  1,  1}0.100 & \cellcolor[rgb]{ 1,  1,  1}0.103 & \cellcolor[rgb]{ 1,  1,  1}0.072 & \cellcolor[rgb]{ 1,  1,  1}0.099 & \cellcolor[rgb]{ 1,  1,  1}0.105 & \cellcolor[rgb]{ 1,  1,  1}0.075 & \cellcolor[rgb]{ 1,  1,  1}0.096 & \cellcolor[rgb]{ 1,  1,  1}0.135 \\
   \hline
    \rowcolor[rgb]{ .886,  .937,  .855} ETS   & \cellcolor[rgb]{ 1,  1,  1}0.070 & \cellcolor[rgb]{ 1,  1,  1}0.099 & \cellcolor[rgb]{ 1,  1,  1}0.116 & \cellcolor[rgb]{ 1,  1,  1}0.071 & \cellcolor[rgb]{ 1,  1,  1}0.097 & \cellcolor[rgb]{ 1,  1,  1}0.119 & \cellcolor[rgb]{ 1,  1,  1}0.079 & \cellcolor[rgb]{ 1,  1,  1}0.092 & \cellcolor[rgb]{ 1,  1,  1}0.154 \\
   \hline
    \rowcolor[rgb]{ .886,  .937,  .855} EXP   & \cellcolor[rgb]{ 1,  1,  1}0.070 & \cellcolor[rgb]{ 1,  1,  1}0.099 & \cellcolor[rgb]{ 1,  1,  1}0.116 & \cellcolor[rgb]{ 1,  1,  1}0.071 & \cellcolor[rgb]{ 1,  1,  1}0.097 & \cellcolor[rgb]{ 1,  1,  1}0.119 & \cellcolor[rgb]{ 1,  1,  1}0.079 & \cellcolor[rgb]{ 1,  1,  1}0.092 & \cellcolor[rgb]{ 1,  1,  1}0.154 \\
   \hline
    \rowcolor[rgb]{ .886,  .937,  .855} EXT   & \cellcolor[rgb]{ 1,  1,  1}0.074 & \cellcolor[rgb]{ 1,  1,  1}0.178 & \cellcolor[rgb]{ 1,  1,  1}0.136 & \cellcolor[rgb]{ 1,  1,  1}0.071 & \cellcolor[rgb]{ 1,  1,  1}0.099 & \cellcolor[rgb]{ 1,  1,  1}0.101 & \cellcolor[rgb]{ 1,  1,  1}0.072 & \cellcolor[rgb]{ 1,  1,  1}0.095 & \cellcolor[rgb]{ 1,  1,  1}0.124 \\
   \hline
    \rowcolor[rgb]{ .886,  .937,  .855} GRB   & \cellcolor[rgb]{ 1,  1,  1}0.071 & \cellcolor[rgb]{ 1,  1,  1}0.169 & \cellcolor[rgb]{ 1,  1,  1}0.109 & \cellcolor[rgb]{ 1,  1,  1}0.188 & \cellcolor[rgb]{ 1,  1,  1}0.098 & \cellcolor[rgb]{ 1,  1,  1}0.111 & \cellcolor[rgb]{ 1,  1,  1}0.076 & \cellcolor[rgb]{ 1,  1,  1}0.096 & \cellcolor[rgb]{ 1,  1,  1}0.192 \\
   \hline
    \rowcolor[rgb]{ .886,  .937,  .855} HUB   & \cellcolor[rgb]{ 1,  1,  1}0.074 & \cellcolor[rgb]{ 1,  .949,  .8}\textcolor[rgb]{ 1,  0,  0}{\textbf{0.095}} & \cellcolor[rgb]{ 1,  1,  1}0.111 & \cellcolor[rgb]{ 1,  1,  1}0.074 & \cellcolor[rgb]{ 1,  .949,  .8}\textcolor[rgb]{ 1,  0,  0}{\textbf{0.094}} & \cellcolor[rgb]{ 1,  1,  1}0.113 & \cellcolor[rgb]{ 1,  1,  1}0.075 & \cellcolor[rgb]{ 1,  1,  1}0.093 & \cellcolor[rgb]{ 1,  1,  1}0.126 \\
    \hline
    \rowcolor[rgb]{ .663,  .816,  .557} KNN   & \multicolumn{1}{r}{\cellcolor[rgb]{ 1,  1,  1}0.070} & \multicolumn{1}{r}{\cellcolor[rgb]{ 1,  1,  1}0.127} & \cellcolor[rgb]{ 1,  1,  1}0.107 & \cellcolor[rgb]{ 1,  0,  0}\textcolor[rgb]{ 1,  1,  1}{\textbf{0.044}} & \multicolumn{1}{r}{\cellcolor[rgb]{ 1,  1,  1}0.115} & \cellcolor[rgb]{ 1,  1,  1}0.103 & \cellcolor[rgb]{ 1,  .949,  .8}\textcolor[rgb]{ 1,  0,  0}{\textbf{0.059}} & \multicolumn{1}{r}{\cellcolor[rgb]{ 1,  1,  1}0.105} & \cellcolor[rgb]{ 1,  1,  1}0.117 \\
    \hline
    \rowcolor[rgb]{ .886,  .937,  .855} LAR   & \cellcolor[rgb]{ 1,  1,  1}0.074 & \cellcolor[rgb]{ 1,  1,  1}0.100 & \cellcolor[rgb]{ 1,  1,  1}0.100 & \cellcolor[rgb]{ 1,  1,  1}0.074 & \cellcolor[rgb]{ 1,  1,  1}0.099 & \cellcolor[rgb]{ 1,  1,  1}0.102 & \cellcolor[rgb]{ 1,  1,  1}0.075 & \cellcolor[rgb]{ 1,  1,  1}0.092 & \cellcolor[rgb]{ 1,  1,  1}0.130 \\
   \hline
    \rowcolor[rgb]{ .886,  .937,  .855} LAS   & \cellcolor[rgb]{ 1,  1,  1}0.071 & \cellcolor[rgb]{ 1,  1,  1}0.100 & \cellcolor[rgb]{ 1,  1,  1}0.110 & \cellcolor[rgb]{ 1,  1,  1}0.071 & \cellcolor[rgb]{ 1,  1,  1}0.099 & \cellcolor[rgb]{ 1,  1,  1}0.112 & \cellcolor[rgb]{ 1,  1,  1}0.076 & \cellcolor[rgb]{ 1,  1,  1}0.096 & \cellcolor[rgb]{ 1,  1,  1}0.145 \\
   \hline
    \rowcolor[rgb]{ .886,  .937,  .855} LGB   & \cellcolor[rgb]{ 1,  1,  1}0.126 & \cellcolor[rgb]{ 1,  1,  1}0.100 & \cellcolor[rgb]{ 1,  1,  1}0.108 & \cellcolor[rgb]{ 1,  1,  1}0.125 & \cellcolor[rgb]{ 1,  1,  1}0.099 & \cellcolor[rgb]{ 1,  1,  1}0.103 & \cellcolor[rgb]{ 1,  1,  1}0.070 & \cellcolor[rgb]{ 1,  1,  1}0.151 & \cellcolor[rgb]{ 1,  1,  1}0.141 \\
   \hline
    \rowcolor[rgb]{ .886,  .937,  .855} LIN   & \cellcolor[rgb]{ 1,  1,  1}0.074 & \cellcolor[rgb]{ 1,  1,  1}0.100 & \cellcolor[rgb]{ 1,  1,  1}0.098 & \cellcolor[rgb]{ 1,  1,  1}0.074 & \cellcolor[rgb]{ 1,  1,  1}0.099 & \cellcolor[rgb]{ 1,  .949,  .8}\textcolor[rgb]{ 1,  0,  0}{\textbf{0.100}} & \cellcolor[rgb]{ 1,  1,  1}0.075 & \cellcolor[rgb]{ 1,  1,  1}0.092 & \cellcolor[rgb]{ 1,  1,  1}0.126 \\
   \hline
    \rowcolor[rgb]{ .886,  .937,  .855} LLA   & \cellcolor[rgb]{ 1,  1,  1}0.074 & \cellcolor[rgb]{ 1,  1,  1}0.099 & \cellcolor[rgb]{ 1,  1,  1}0.100 & \cellcolor[rgb]{ 1,  1,  1}0.074 & \cellcolor[rgb]{ 1,  1,  1}0.098 & \cellcolor[rgb]{ 1,  1,  1}0.102 & \cellcolor[rgb]{ 1,  1,  1}0.075 & \cellcolor[rgb]{ 1,  1,  1}0.094 & \cellcolor[rgb]{ 1,  1,  1}0.130 \\
   \hline
    \rowcolor[rgb]{ .886,  .937,  .855} NAV   & \cellcolor[rgb]{ 1,  1,  1}0.446 & \cellcolor[rgb]{ 1,  .78,  .808}\textcolor[rgb]{ .612,  0,  .024}{0.227} & \cellcolor[rgb]{ 1,  .922,  .612}\textcolor[rgb]{ .612,  .341,  0}{\textbf{1.131}} & \cellcolor[rgb]{ 1,  1,  1}0.446 & \cellcolor[rgb]{ 1,  1,  1}0.227 & \cellcolor[rgb]{ 1,  1,  1}1.131 & \cellcolor[rgb]{ 1,  1,  1}0.446 & \cellcolor[rgb]{ 1,  1,  1}0.227 & \cellcolor[rgb]{ 1,  1,  1}1.131 \\
   \hline
    \rowcolor[rgb]{ .886,  .937,  .855} OMP   & \cellcolor[rgb]{ 1,  1,  1}0.072 & \cellcolor[rgb]{ 1,  1,  1}0.101 & \cellcolor[rgb]{ 1,  1,  1}0.100 & \cellcolor[rgb]{ 1,  1,  1}0.072 & \cellcolor[rgb]{ 1,  1,  1}0.099 & \cellcolor[rgb]{ 1,  1,  1}0.102 & \cellcolor[rgb]{ 1,  1,  1}0.074 & \cellcolor[rgb]{ 1,  .949,  .8}\textcolor[rgb]{ 1,  0,  0}{\textbf{0.090}} & \cellcolor[rgb]{ 1,  1,  1}0.130 \\
   \hline
    \rowcolor[rgb]{ .886,  .937,  .855} POL   & \cellcolor[rgb]{ 1,  1,  1}0.086 & \cellcolor[rgb]{ 1,  1,  1}0.099 & \cellcolor[rgb]{ 1,  1,  1}0.116 & \cellcolor[rgb]{ 1,  1,  1}0.087 & \cellcolor[rgb]{ 1,  1,  1}0.097 & \cellcolor[rgb]{ 1,  1,  1}0.119 & \cellcolor[rgb]{ 1,  1,  1}0.098 & \cellcolor[rgb]{ 1,  1,  1}0.092 & \cellcolor[rgb]{ 1,  1,  1}0.154 \\
   \hline
    \rowcolor[rgb]{ .886,  .937,  .855} PRO   & \cellcolor[rgb]{ 1,  1,  1}0.088 & \cellcolor[rgb]{ 1,  1,  1}0.147 & \cellcolor[rgb]{ 1,  1,  1}0.106 & \cellcolor[rgb]{ 1,  1,  1}0.089 & \cellcolor[rgb]{ 1,  1,  1}0.145 & \cellcolor[rgb]{ 1,  1,  1}0.109 & \cellcolor[rgb]{ 1,  1,  1}0.081 & \cellcolor[rgb]{ 1,  1,  1}0.100 & \cellcolor[rgb]{ 1,  1,  1}0.106 \\
   \hline
    \rowcolor[rgb]{ .886,  .937,  .855} RAN   & \cellcolor[rgb]{ 1,  1,  1}0.071 & \cellcolor[rgb]{ 1,  1,  1}0.098 & \cellcolor[rgb]{ 1,  1,  1}0.111 & \cellcolor[rgb]{ 1,  1,  1}0.071 & \cellcolor[rgb]{ 1,  1,  1}0.158 & \cellcolor[rgb]{ 1,  1,  1}0.139 & \cellcolor[rgb]{ 1,  1,  1}0.077 & \cellcolor[rgb]{ 1,  1,  1}0.097 & \cellcolor[rgb]{ 1,  1,  1}0.147 \\
   \hline
    \rowcolor[rgb]{ .886,  .937,  .855} RID   & \cellcolor[rgb]{ 1,  1,  1}0.074 & \cellcolor[rgb]{ 1,  1,  1}0.099 & \cellcolor[rgb]{ 1,  1,  1}0.100 & \cellcolor[rgb]{ 1,  1,  1}0.074 & \cellcolor[rgb]{ 1,  1,  1}0.098 & \cellcolor[rgb]{ 1,  1,  1}0.102 & \cellcolor[rgb]{ 1,  1,  1}0.075 & \cellcolor[rgb]{ 1,  1,  1}0.093 & \cellcolor[rgb]{ 1,  1,  1}0.130 \\
   \hline
    \rowcolor[rgb]{ .886,  .937,  .855} TBA   & \cellcolor[rgb]{ 1,  1,  1}0.077 & \cellcolor[rgb]{ 1,  1,  1}0.122 & \cellcolor[rgb]{ 1,  1,  1}0.106 & \cellcolor[rgb]{ 1,  1,  1}0.077 & \cellcolor[rgb]{ 1,  1,  1}0.135 & \cellcolor[rgb]{ 1,  1,  1}0.106 & \cellcolor[rgb]{ 1,  1,  1}0.106 & \cellcolor[rgb]{ 1,  1,  1}0.102 & \cellcolor[rgb]{ 1,  .949,  .8}\textcolor[rgb]{ 1,  0,  0}{\textbf{0.090}} \\
   \hline
    \rowcolor[rgb]{ .886,  .937,  .855} THE   & \cellcolor[rgb]{ 1,  1,  1}0.070 & \cellcolor[rgb]{ 1,  1,  1}0.100 & \cellcolor[rgb]{ 1,  1,  1}0.115 & \cellcolor[rgb]{ 1,  1,  1}0.071 & \cellcolor[rgb]{ 1,  1,  1}0.098 & \cellcolor[rgb]{ 1,  1,  1}0.117 & \cellcolor[rgb]{ 1,  1,  1}0.079 & \cellcolor[rgb]{ 1,  1,  1}0.092 & \cellcolor[rgb]{ 1,  1,  1}0.150 \\
   \hline
    \rowcolor[rgb]{ .886,  .937,  .855} XGB   & \cellcolor[rgb]{ 1,  1,  1}0.077 & \cellcolor[rgb]{ 1,  1,  1}0.130 & \cellcolor[rgb]{ 1,  1,  1}0.307 & \cellcolor[rgb]{ 1,  1,  1}0.131 & \cellcolor[rgb]{ 1,  1,  1}0.096 & \cellcolor[rgb]{ 1,  1,  1}0.157 & \cellcolor[rgb]{ 1,  1,  1}0.081 & \cellcolor[rgb]{ 1,  1,  1}0.109 & \cellcolor[rgb]{ 1,  1,  1}0.182 \\
    \hline
    \end{tabular}%
  \label{tab:RMSE-D1}%
  \end{adjustbox}
\end{table}%

\clearpage

\section*{Declarations}
\begin{itemize}
\item Acknowledgements: The authors sincerely thank Prof. K.S. Gandhi, Department of Chemical Engineering, Indian Institute of Science, Bangalore, for insightful discussions and critical comments on the manuscript.
\item Funding: The authors declare that no funds, grants, or other support were received during the preparation of this manuscript.
\item Conflict of interest/Competing interests: The authors have no competing interests to declare that are relevant to the content of this article.
\item Ethics approval: The research meets all ethical guidelines, including adherence to the legal requirements of the study country.
\item Consent to participate: Not applicable.
\item Consent for publication: All authors agreed with the content and that all gave explicit consent to submit and that they obtained consent from the responsible authorities at the Indian Institute of Science, where the work has been carried out, before the work is submitted.
\item Availability of data and materials: The Global Mean Temperature data was obtained from NASA's website (https://data.giss.nasa.gov/gistemp/). The datasets generated during and/or analysed during the current study are available from the corresponding author, Dr D. Niyogi, on reasonable request.
\item Code availability:  Available from the corresponding author, Dr D. Niyogi, upon reasonable request.
\item Authors' contributions: Prof. J. Srinivasan contributed to the study conception. Material preparation, data collection and analysis were performed by Dr Debdarsan Niyogi. The first draft of the manuscript was written by Dr Debdarsan Niyogi and all authors commented on previous versions of the manuscript. All authors read and approved the final manuscript.
\end{itemize}

\clearpage


\bibliographystyle{unsrtnat}
\bibliography{gmtvar.bib}

\end{document}